\documentclass[11pt]{article}
\usepackage{graphics}
\usepackage{hyperref}
\usepackage{amsmath}
\usepackage{amssymb}
\usepackage{color}

\title{\bf Lectures on \\
perturbative unitarity and decoupling in Higgs physics\footnote{Comments are welcome and will inform future versions of these lecture notes.  I hereby release all the figures in these lectures into the public domain.}}

\author{Heather E.\ Logan\thanks{\tt logan@physics.carleton.ca} \\
{\it Ottawa-Carleton Institute for Physics, Carleton University,}\\ 
{\it 1125 Colonel By Drive, Ottawa, Ontario K1S 5B6 Canada}}

\date{July 3, 2022}

\begin{document}

\maketitle

\begin{abstract} 
\noindent
These lectures are a pedagogical introduction to the application of perturbative unitarity to Higgs physics within and beyond the Standard Model (SM).  I begin with a review of how perturbative unitarity arises from quantum mechanical scattering theory and apply it to the classic problem of longitudinal vector boson scattering in the SM to derive the famous upper bound on the Higgs boson mass.  I then consider extended Higgs sectors, using the two-Higgs-doublet model, the scalar septet model, and the Georgi-Machacek model as case studies.  I discuss the resulting Higgs coupling sum rules as well as the intertwined constraints on masses and couplings of the additional Higgs bosons in these models, highlighting the connection between perturbative unitarity and the decoupling limit.  I finish with a direct review of the decoupling limit in the two-Higgs-doublet model and a digression on the alignment limit.
These notes are based on lectures delivered over the past few years in a graduate-level course on beyond-the-SM phenomenology and are meant as a companion volume to my TASI 2013 lectures on Higgs physics.
\end{abstract}

\newpage
\tableofcontents
\newpage

\section{Introduction}

Perturbative unitarity is one of the most powerful tools available for understanding the theoretical structure of spontaneously broken gauge theories.  Its most famous application was the calculation, in 1977, of the upper bound on the mass of the Standard Model (SM) Higgs boson (or equivalently, on the energy scale at which the dynamics of electroweak symmetry breaking must appear) of about 1~TeV by Lee, Quigg, and Thacker~\cite{Lee:1977yc,Lee:1977eg}.\footnote{I really encourage you to read these two papers; they are truly beautiful.}  This calculation dictated the technical requirements of the CERN Large Hadron Collider (LHC) and dominated the conceptual framework of the hunt for the dynamics of electroweak symmetry breaking up to the discovery of the SM-like Higgs boson in 2012~\cite{Hdiscovery}. 

In these lectures (meant as a companion volume to my TASI 2013 lectures on Higgs physics~\cite{Logan:2014jla}) I give a pedagogical introduction to the use of perturbative unitarity in the context of Higgs physics, both within and beyond the SM.\footnote{Since this is not a review article, I do not make any attempt to provide a comprehensive set of references.  Sorry.}
This document is organized as follows.  In Sec.~\ref{sec:background} I give a brief review of partial wave unitarity in quantum mechanics and discuss its application to perturbative amplitudes in quantum field theory.  In Sec.~\ref{sec:LVBS} I review the calculation of longitudinal vector boson scattering in the SM and show how the famous upper bound on the Higgs mass is obtained.  In Sec.~\ref{sec:2HDM} I extend the calculation to the two-Higgs-doublet model and derive the associated coupling sum rule and the implications for the decoupling behaviour of the model.  In Secs.~\ref{sec:septet} and \ref{sec:GM} I do the same for the scalar septet model and the Georgi-Machacek model, respectively, thereby introducing the complication of singly- and doubly-charged Higgs bosons that contribute to vector boson scattering.  Finally in Sec.~\ref{sec:decoup} I review some details of decoupling and alignment in the two-Higgs-doublet model.
Some concluding remarks are given in Sec.~\ref{sec:summary}, followed by a collection of homework questions in Sec.~\ref{sec:homework}.

\section{Unitarity and perturbative unitarity}
\label{sec:background}

The main goal of these lectures is to study perturbative unitarity in longitudinal vector boson scattering.  This process yields extremely important theoretical constraints on Higgs physics in the SM and beyond, in particular:
\begin{itemize}
\item the SM Higgs mass upper bound;
\item sum rules for Higgs-Vector-Vector couplings beyond the SM; and
\item bounds on ``exotic'' (i.e., higher isospin) Higgs vacuum expectation values (vevs) at high mass.
\end{itemize}

Before jumping in, we will start with a review of the unitarity bound on partial wave amplitudes, which you may remember from the study of scattering theory in quantum mechanics.

\subsection{A quick review of scattering theory}

The unitarity bound on scattering amplitudes is mathematically rigorous and comes directly from the conservation of probability in quantum mechanics.\footnote{Notice the lack of the word ``perturbative'' in the preceding sentence.  The {\it perturbative} unitarity bound is less mathematically sharp because there is no clear line at which a theory goes from being perturbative to being nonperturbative.  More on this later.}\footnote{The derivation in this subsection follows Chapter 19 of Gasiorowicz 3rd edition~\cite{Gasiorowicz}; you can find a similar derivation in Chapter 10 of Griffiths 3rd edition~\cite{Griffiths}.  The same result can also be derived using the Optical Theorem; see Sec.\ 24.1.5 of Schwartz's textbook on Quantum Field Theory~\cite{Schwartz}.}
Start with an incoming plane wave propagating in the $z$ direction.  This initial wavefunction can be written as 
\begin{equation}
	\psi(\vec r) =  e^{i \vec k \cdot \vec r} = e^{i k r \cos \theta},
\end{equation}
where we used $\vec k = k \hat z$ and $z = r \cos \theta$.  This plane wave can be rewritten in terms of spherical Bessel functions $j_\ell$ and Legendre polynomials $P_\ell$ as
\begin{equation}
	\psi(\vec r) = \sum_{\ell = 0}^{\infty} (2 \ell + 1) (i)^\ell j_\ell(kr) P_\ell(\cos\theta).
\end{equation}
For $kr \gg \ell$, i.e., at distances a large number of wavelengths from the scattering centre (with ``large'' being large compared to the angular momentum quantum number $\ell$), this can be approximated by the expression
\begin{equation}
	\psi(\vec r) \simeq \frac{i}{2k} \sum_{\ell = 0}^{\infty} (2 \ell + 1) (i)^\ell 
	\left[ \frac{e^{-i (kr - \ell \pi/2)}}{r} - \frac{e^{+i (kr - \ell \pi/2)}}{r} \right] P_\ell (\cos\theta),
\end{equation}
where the first term in the square brackets is an incoming spherical wave and the second term is an outgoing spherical wave (you can prove this to yourself by computing the probability current density for each of these two waves separately).

This is still an {\it undisturbed} plane wave.  Because of causality, the only thing that the scattering centre can do is to change the phase (and possibly the amplitude) of the {\it outgoing} spherical wave:
\begin{equation}
	\psi(\vec r) \simeq \frac{i}{2k} \sum_{\ell = 0}^{\infty} (2 \ell + 1) (i)^\ell 
	\left[ \frac{e^{-i (kr - \ell \pi/2)}}{r} - S_\ell(k) \frac{e^{+i (kr - \ell \pi/2)}}{r} \right] P_\ell (\cos\theta),
\end{equation}
where again the approximation is good when $kr \gg \ell$.  In order for the incoming probability flux to equal the outgoing probability flux, we must have $|S_{\ell}(k)| = 1$, i.e., $S_{\ell}(k)$ is purely a phase.  If there is absorption of some of the probability flux (or, equivalently, conversion of some of the flux into a different type of particle), then we can have $|S_{\ell}(k)| \leq 1$.  This upper bound on the magnitude of $S_{\ell}(k)$ is going to be the source of the unitarity bound on the partial wave amplitudes.

\begin{figure}
\center
\resizebox{0.4\textwidth}{!}{\includegraphics{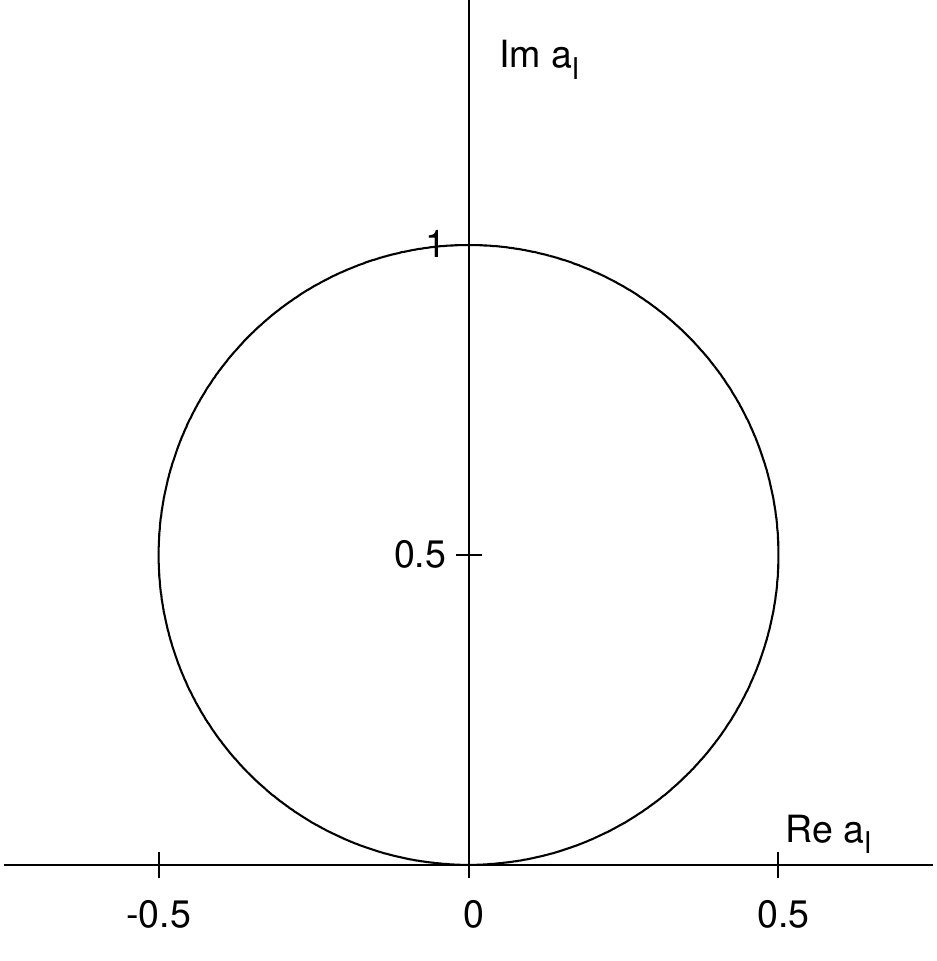}}
\caption{Sketch of the allowed range of each partial wave amplitude $a_\ell$ in the complex plane.  Unitarity constrains the value of each $a_\ell$ to lie on or within the circle.}
\label{fig:al-complex-plane}
\end{figure}

Now we rearrange our expression for the wavefunction to separate back out the original incoming plane wave (and do a bit of algebra):
\begin{equation}
	\psi(\vec r) \simeq e^{ikr \cos\theta} 
	+ \left[ \sum_{\ell = 0}^{\infty} (2 \ell + 1) \frac{S_\ell(k) - 1}{2 i k} P_\ell(\cos\theta) \right]
	\frac{e^{ikr}}{r}. 
\end{equation}
The first term is the original plane wave.  The $e^{ikr}/r$ is a pure outgoing spherical wave.  The stuff in the square brackets multiplying the outgoing spherical wave contains the $\theta$ dependence and all of the coefficients that determine the strength of the scattering, and we will call it $f(\theta)$ for short:
\begin{equation}
	f(\theta) = \frac{1}{k} \sum_{\ell = 0}^{\infty} (2 \ell + 1) \frac{S_{\ell}(k) - 1}{2 i} P_\ell(\cos\theta) 
	\equiv \frac{1}{k} \sum_{\ell = 0}^{\infty} (2 \ell + 1) a_\ell(k) P_\ell(\cos\theta),
\end{equation}
where we have defined $a_\ell(k) = (S_\ell(k) - 1)/2i$.  This complex quantity $a_\ell(k)$ is the \emph{partial wave amplitude} for the $\ell$th partial wave (the explicit $k$ dependence just serves to remind us that the scattering probability distribution can depend on the wavenumber).  Using the fact that $|S_\ell(k)| \leq 1$, we can see that $a_\ell(k)$ must live on or within a circle in the complex plane of radius $1/2$ and centred at $({\rm Re} \, a_\ell, {\rm Im} \, a_\ell) = (0,1/2)$, as sketched in Fig.~\ref{fig:al-complex-plane}.

This is the \emph{unitarity bound} on each of the partial wave amplitudes $a_\ell$, and it arises solely from the requirement that the outgoing flux cannot be greater than the incoming flux.  The unitarity bound implies that the following constraints on $a_\ell$ must be satisfied:
\begin{equation}
	| a_\ell | \leq 1; \qquad | {\rm Re} \, a_\ell | \leq 1/2; \qquad 0 \leq {\rm Im} \, a_\ell \leq 1.
\end{equation}

\subsection{Translation to quantum field theory}

Our next task is to translate this bound into the language of the matrix element $\mathcal{M}$ that we compute in quantum field theory.  This is easily done by computing the differential cross section for a scattering process in each of the two formalisms and comparing them.  In quantum mechanics we have,
\begin{equation}
	\frac{d \sigma}{d \Omega} = | f(\theta) |^2 
		= \left| \frac{1}{k} \sum_{\ell = 0}^{\infty} (2 \ell + 1) a_\ell(k) P_\ell(\cos\theta) \right|^2,
\end{equation}
where $d \Omega$ is the differential solid angle.  When the scattering of a single particle off a fixed potential is reinterpreted as the scattering of two particles in the centre-of-momentum (CM) frame, $k$ becomes the momentum $|\vec p|$ of \emph{one} of the particles in that frame.   

Meanwhile, the differential cross section in quantum field theory for two-to-two scattering is
\begin{equation}
	\left( \frac{d \sigma}{d \Omega} \right)_{CM} = \frac{ | \mathcal{M} |^2 }{ 64 \pi^2 E_{CM}^2 },
\end{equation}
where $E_{CM}$ is the \emph{total} energy in the CM frame and I have taken all masses equal for simplicity.\footnote{The same formula holds for unequal masses when these masses can be neglected compared to the scattering energy.}  We therefore identify
\begin{equation}
	\mathcal M = 8 \pi \frac{E_{CM}}{|\vec p|} \sum_{\ell = 0}^{\infty} (2 \ell + 1) a_\ell (|\vec p|) P_\ell(\cos\theta).
\end{equation}
When $E_{CM}$ is much larger than any of the particle masses (which will be the case in most practical applications of perturbative unitarity constraints), $E_{CM} \approx 2 |\vec p|$, so in this limit we have
\begin{equation}
\boxed{
	\mathcal M = 16 \pi \sum_{\ell = 0}^{\infty} (2 \ell + 1) a_\ell (|\vec p|) P_\ell(\cos\theta),
}
\label{eq:Mformula}
\end{equation}
where each $a_\ell$ is bounded to live on or within the circle in the complex plane shown in Fig.~\ref{fig:al-complex-plane}.

\subsection{Where ``perturbative'' comes in}

The bound on the (in general complex) matrix element $\mathcal{M}$ that comes from combining Eq.~(\ref{eq:Mformula}) with the unitarity limits on each $a_\ell$ shown in Fig.~\ref{fig:al-complex-plane} arises from the basic structure of quantum field theory and \emph{must hold in any complete model}.\footnote{Incomplete models, such as effective theories containing nonrenormalizable operators (four-Fermi theory being the classic example) or the SM with no Higgs boson (which we will study in the next section), are pretty much guaranteed to violate perturbative unitarity at high enough scattering energy.  This is a sign that new physics will come in at or below that energy scale to ``fix'' the unitarity-violating behaviour, even if that new physics is just the new resonances of a strongly-coupled theory.  More on this in the next section.}  This raises the question of how these constraints could possibly be useful for constraining parameters of a model when they are guaranteed to hold anyway.

The answer is in the word \emph{perturbative}.  While the full, nonperturbative result for $\mathcal{M}$ must automatically satisfy the unitarity bound, the perturbative calculation of $\mathcal{M}$ to any particular order in perturbation theory need not.  If the perturbative (say, tree-level) calculation of $\mathcal{M}$ violates the unitarity bound, it must be the case that the higher-order contributions to $\mathcal{M}$ have to be sufficient to ``save the day'' and yield a final result that satisfies unitarity (sometimes this is phrased as the higher-order corrections \emph{restoring unitarity}).  This in turn implies that the higher-order contributions to $\mathcal{M}$ are important enough that they can't really be neglected, i.e., that our perturbative calculation is not a good approximation for $\mathcal{M}$.  In this situation we say that perturbation theory has broken down; this usually happens because the coupling parameter in which we're making the perturbative expansion is too large.

Notice in fact that tree-level matrix elements are purely real; a purely real nonzero $a_\ell$ technically already violates unitarity, because the disc of allowed values in Fig.~\ref{fig:al-complex-plane} only touches the real axis at the origin!  This is not considered to be a serious problem, because near the origin its bounding circle can be approximated by a quadratic, so that the imaginary part is of the right size to be generated by terms of the next order in perturbation theory (i.e., one-loop diagrams).  In practical calculations, the bound is usually taken to be 
\begin{equation}
\boxed{
	| a_\ell | < 1 \quad {\rm or} \quad | {\rm Re} \, a_\ell | < 1/2.
}
\end{equation}  
I personally prefer to use the latter when applying perturbative unitarity bounds to tree-level matrix elements, because the tree-level matrix element is real anyway and because it keeps the theory a little bit farther away from the nonperturbative danger zone (which is important if one wants to be able to trust phenomenological predictions that are calculated perturbatively).\footnote{On the other hand, there are good arguments to use the former, looser, bound if one is using the result to decide how big of a tunnel you need to dig for your next collider!}  This is largely a matter of taste and you will encounter both choices in the literature.

\section{Longitudinal vector boson scattering and the SM Higgs}
\label{sec:LVBS}

The classic calculation of Lee, Quigg, and Thacker~\cite{Lee:1977yc,Lee:1977eg} considers two-to-two scattering of massive vector bosons (i.e., $W$ and $Z$ bosons) in the SM.  The interesting physics (for our purposes) comes from the scattering of the longitudinal polarization states of the $W$ and $Z$ bosons; i.e., the polarization states that are present only because of electroweak symmetry breaking.  For concreteness, we'll first consider $W^+ W^- \to W^+ W^-$.

\subsection{Four-point diagram}

\begin{figure} 
\center
\includegraphics{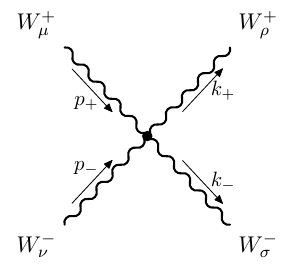} 
\caption{Four-point diagram contributing to $W^+ W^- \to W^+ W^-$ scattering.  The indicated $W$ boson charges are defined to be flowing in the same direction as the corresponding momentum arrows.}
\label{fig:WW4pt}
\end{figure}

We start with the four-point diagram shown in Fig.~\ref{fig:WW4pt}.  The Feynman rule for the vertex is\footnote{For a complete and internally self-consistent compilation of the Feynman rules of the SM, in all possible conventions, see Ref.~\cite{Romao:2012pq}.}
\begin{equation}
	i g^2 ( 2 g_{\mu \sigma} g_{\nu \rho} - g_{\mu \nu} g_{\rho \sigma} 
		- g_{\mu \rho} g_{\nu \sigma}).
\end{equation}
Computing the matrix element just involves contracting this vertex with the polarization vectors of the four external gauge bosons.  For \emph{transverse} gauge boson polarization vectors, the matrix element from this diagram is independent of scattering energy, and $g$ is small enough that it satisfies the perturbative unitarity constraint.  For \emph{longitudinal} polarizations, however, the amplitude coming from this diagram grows with energy!  Let's compute it.

Let the four-momenta of the initial and final $W$ bosons be
\begin{align}
	p_+ &= (E, 0, 0, p)  &  k_+ &= (E, p \sin\theta, 0, p \cos\theta) \nonumber \\
	p_- &= (E, 0, 0, -p)  &  k_- &= (E, -p \sin\theta, 0, -p \cos\theta).
\end{align}
Then the longitudinal polarization vectors for each of the external $W$ bosons are\footnote{These can be derived by starting from a $W$ boson at rest with four-momentum $k = (M_W, 0, 0, 0)$ and polarization four-vector $\epsilon = (0,0,0,1)$ and then applying the appropriate relativistic boost along the $z$ direction; see section 21.2 of Peskin and Schroeder~\cite{Peskin}.}
\begin{align}
	\epsilon_L(p_+) &= \left( \frac{p}{M_W}, 0, 0, \frac{E}{M_W} \right)  
		&  \epsilon_L(k_+) &= \left( \frac{p}{M_W}, \frac{E}{M_W} \sin\theta, 0, 
			\frac{E}{M_W} \cos\theta \right) \nonumber \\
	\epsilon_L(p_-) &= \left( \frac{p}{M_W}, 0, 0, -\frac{E}{M_W} \right)
		&  \epsilon_L(k_-) &= \left( \frac{p}{M_W}, -\frac{E}{M_W} \sin\theta, 0,
			-\frac{E}{M_W} \cos\theta \right).
\end{align}
Note that $\epsilon_L(p_i) \cdot p_i = 0$ and $\left( \epsilon_L(p_i) \right)^2 = -1$,\footnote{I use the metric $g^{\mu \nu} = {\rm diag}(1, -1, -1, -1)$, in which an on-shell particle has $p^2 = m^2$.} as they should be.

At high energies $E \gg M_W$, the matrix element from this four-point diagram ends up taking the form
\begin{equation}
	i \mathcal{M} = i g^2 \frac{E^4}{M_W^4} \left[ \cdots \right].
\end{equation}
In particular, it grows with energy like $E^4$, and is thus guaranteed to violate perturbative unitarity at high energies!  Fortunately we are not done; there are more diagrams to compute.

\subsection{Diagrams with $s$- and $t$-channel gauge bosons}

\begin{figure}
\center
\includegraphics{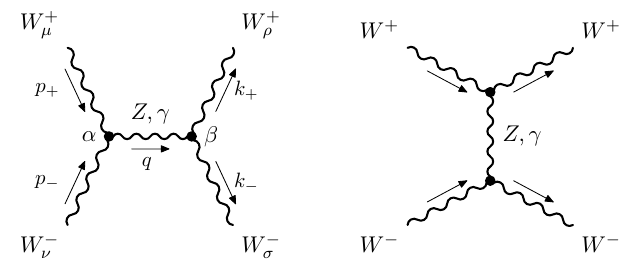} 
\caption{Diagrams with $s$- and $t$-channel gauge bosons contributing to $W^+ W^- \to W^+ W^-$ scattering.}
\label{fig:WWstgauge}
\end{figure}

Next let's tackle the diagrams with $s$- and $t$-channel gauge boson exchange,\footnote{There is no $u$-channel diagram for this particular scattering process because it would require exchange of a doubly-charged gauge boson, which does not exist in the SM.} which are shown in Fig.~\ref{fig:WWstgauge}.  Charge conservation implies that the exchanged gauge bosons must be neutral; i.e., the $Z$ boson and the photon both contribute.

The Feynman rule for the three-gauge-boson vertex is complicated and depends on the directions of the momenta of the three bosons involved.  For the vertex labeled $\alpha$ in the $s$-channel diagram in Fig.~\ref{fig:WWstgauge}, the Feynman rule is (we use the Lagrangian sign conventions of Ref.~\cite{Peskin}),
\begin{equation}
	i \left\{ \begin{array}{c} e \\ g \cos\theta_W \end{array} \right\} 
		\left[ g_{\mu \nu} (p_- - p_+)_{\alpha} + g_{\mu \alpha} (q + p_+)_{\nu}
			+ g_{\nu \alpha} (-q - p_-)_{\mu} \right],
\end{equation}
where the $e$ is the coupling for the photon and the $g \cos\theta_W$ is the coupling for the $Z$ boson.  Meanwhile the $Z$ and photon propagators, in unitarity gauge, are, respectively,
\begin{equation}
	\frac{-i}{q^2 - M_Z^2} \left[ g_{\alpha \beta} - \frac{q_{\alpha} q_{\beta}}{M_Z^2} \right] \qquad
	{\rm and} \qquad \frac{-i g_{\alpha \beta}}{q^2}.
\end{equation}

Each of the four diagrams shown in Fig.~\ref{fig:WWstgauge} individually grows with energy like $E^4/M_W^4$ (technically there are also terms with powers of energy as high as $E^6/M_W^4 M_Z^2$, but these cancel within each diagram).  After some tedious calculating, the $E^4/M_W^4$ pieces \emph{cancel} between the four-point diagram of the previous subsection and the four diagrams considered here!  This is an elegant consequence of the relationships among the four-point and three-point couplings that arise from the structure of the underlying SU(2) gauge theory.

\subsection{Perturbative unitarity bound without the Higgs boson}

The cancellation of the terms proportional to $E^4$ in the sum of the four-point diagram and the diagrams involving $s$- and $t$-channel gauge bosons leaves terms proportional to $E^2/M_W^2$.  For scattering of longitudinally polarized $W$ bosons, $W^+_L W^-_L \to W^+_L W^-_L$, after combining the above diagrams we have
\begin{equation}
	i \mathcal{M} = i \frac{g^2}{4 M_W^2} \left[s + t + \mathcal{O}(M_W^2) \right],
\end{equation}
where $s$, $t$, and $u$ are the usual Mandelstam variables\footnote{In terms of the four-momenta defined in Fig.~\ref{fig:WW4pt}, the Mandelstam variables are defined as $s = (p_+ + p_-)^2 = (k_+ + k_-)^2$, $t = (p_+ - k_+)^2 = (k_- - p_-)^2$, and $u = (p_+ - k_-)^2 = (k_+ - p_-)^2$.} and are proportional to $E^2$.  We will ignore the $\mathcal{O}(M_W^2)$ piece in the square brackets because it will turn out to give a contribution to the full amplitude of order $g^2$, which is numerically small compared to the perturbative unitarity bound and, of course, does not grow with energy.  Using $M_W^2 = g^2 v^2/4$ (where $v \simeq 246$~GeV is the SM Higgs vev), we can rewrite the amplitude so far as
\begin{equation}
	i \mathcal{M} \simeq i \frac{1}{v^2} (s + t),
	\label{eq:Mgaugepart}
\end{equation}
where I'm now just ignoring the contributions of order $g^2$.

Notice that the calculation up to now includes only the physical degrees of freedom that had been produced and detected on-shell in experiments prior to the Higgs boson discovery of 2012.  The fact that the amplitude grows with increasing energy tells us that something is missing, which is not surprising because we have not yet included the SM diagrams involving $s$- and $t$-channel Higgs boson exchange.  This is, in fact, the point.  Let's proceed to apply the perturbative unitarity bound to this ``Higgsless'' amplitude and estimate the scattering energy at which this process violates perturbative unitarity.  We have,
\begin{eqnarray}
	\mathcal{M} &=& 16 \pi \sum_{\ell = 0}^{\infty} (2 \ell + 1) a_\ell P_\ell(\cos\theta) \nonumber \\
	&=& 16 \pi a_0 + \cdots.
\end{eqnarray}
Ignoring for now the $t$ dependence (since $t$ depends on $\cos\theta$) and applying the bound $|{\rm Re} \, a_0| \leq 1/2$ gives
\begin{equation}
	a_0 \sim \frac{s}{16 \pi v^2} \leq 1/2 \qquad \Rightarrow \qquad 
	s \lesssim 8 \pi v^2 \simeq (1200~{\rm GeV})^2.
\end{equation}
In other words, this theory (without the Higgs boson) \emph{breaks down} at the TeV scale!

Remember that quantum field theory must conserve probability no matter what.  Generally what this violation of perturbative unitarity means is that (a) New Physics appears below this scale to fix things (in the SM, this ``new physics'' is just the Higgs boson), or (b) the theory becomes strongly coupled so that higher-order diagrams become large enough to fix unitarity (this possibility would have resulted in so-called strong $WW$ scattering at the TeV scale, a possibility that the LHC was equipped to study).

\subsection{Diagrams with SM Higgs boson exchange}

\begin{figure}
\center
\includegraphics{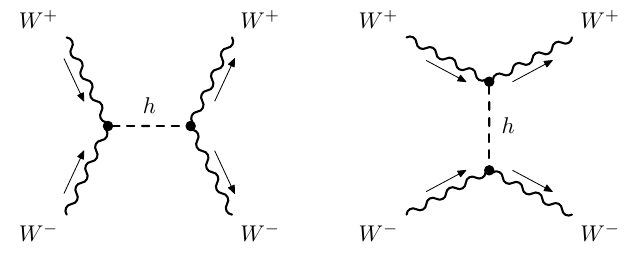} 
\caption{Diagrams with $s$- and $t$-channel SM Higgs bosons contributing to $W^+ W^- \to W^+ W^-$ scattering.}
\label{fig:WWstHiggs}
\end{figure}

We now add in the rest of the diagrams that enter the SM calculation of longitudinal $WW$ scattering, which involve $s$- and $t$-channel Higgs boson exchange and are shown in Fig.~\ref{fig:WWstHiggs}.  The Feynman rule for each of the vertices is
\begin{equation}
	\frac{2 i M_W^2}{v} g_{\mu\nu} = i \frac{g^2 v}{2} g_{\mu\nu}.
\end{equation}
Calculating the two diagrams yields a contribution to the matrix element,
\begin{equation}
	i \mathcal{M} = -i \frac{1}{v^2} \left( \frac{s^2}{s - m_h^2} + \frac{t^2}{t - m_h^2} \right),
	\label{eq:Mst1}
\end{equation}
where $m_h$ is the Higgs mass, the $s^2$ and $t^2$ in the numerators come from the dot products of the $W$ boson longitudinal polarization four-vectors (caused by the $g_{\mu\nu}$ factor in the vertices), and we have not yet made any approximations.  This can be simplified by performing a Taylor expansion of the denominators for $s, t \gg m_h^2$ (i.e., high energy scattering), which gives
\begin{equation}
	i \mathcal{M} \simeq -i \frac{1}{v^2} \left( s + t + 2 m_h^2 + \cdots \right),
	\label{eq:Mhiggspart}
\end{equation}
where we have dropped terms that fall with increasing energy.

Adding the Higgs-exchange contribution in Eq.~(\ref{eq:Mhiggspart}) to the sum of the ``Higgsless'' diagrams in Eq.~(\ref{eq:Mgaugepart}), we see that the $(s + t)$ part \emph{exactly cancels}, leaving a matrix element that does not grow with energy!  (This is actually guaranteed to happen in a renormalizable theory, so it should not be too much of a surprise.)  The left-over ``finite'' piece (meaning, the piece that does not grow with energy) in the high-energy limit is (again neglecting the pure-gauge contribution to $\mathcal{M}$ of order $g^2$),
\begin{equation}
	\mathcal{M} \simeq - \frac{1}{v^2} (2 m_h^2) = 16 \pi a_0 \qquad {\rm or} \qquad
	a_0 \simeq - \frac{1}{16 \pi v^2} \cdot 2 m_h^2.
\end{equation}
Applying the unitarity bound $|{\rm Re} \, a_0| \leq 1/2$ yields an \emph{upper bound on the Higgs boson mass} of 
\begin{equation}
	m_h^2 \leq 4 \pi v^2 \simeq (870~{\rm GeV})^2.
\end{equation}

This is the upper bound on the Higgs boson mass from perturbative unitarity (i.e., we only calculated the tree-level amplitude) of the zeroth partial wave amplitude for longitudinal $W^+W^-$ boson scattering.

\subsection{Coupled-channel analysis}

The upper bound on the Higgs mass that we just found can be refined a bit (read: made a little bit more stringent) by using a \emph{coupled-channel analysis}.  We considered this process:
\begin{equation}
	W^+_L W^-_L \to W^+_L W^-_L.
\end{equation}
The same initial state can also give
\begin{eqnarray}
	&& W^+_L W^-_L \to Z_L Z_L \\
	&& W^+_L W^-_L \to hh,
\end{eqnarray}
and there are also the related processes
\begin{eqnarray}
	Z_L Z_L &\to& W^+_L W^-_L, \ Z_L Z_L, \ hh \\
	hh &\to& W^+_L W^-_L, \ Z_L Z_L, \ hh.
\end{eqnarray}
Each of these tree-level processes gives a finite amplitude proportional to $m_h^2/v^2$ at high scattering energy.

The idea of the coupled-channel analysis is as follows.  Say we start from a quantum mechanical linear combination (i.e., a superposition) of these initial states and want to compute the amplitude to scatter into some linear combination of the final states.  Perturbative unitarity will still require us to apply the bound $|{\rm Re} \, a_0| \leq 1/2$ to the \emph{largest} amplitude that we can achieve in this way.  In practice, we can implement this by computing all of the amplitudes for the processes listed above, assembling them into a matrix, finding the largest eigenvalue, and imposing $|{\rm Re} \, a_0| \leq 1/2$ on that eigenvalue.

A subtlety that arises in this calculation is that for each pair of identical particles (i.e., $Z_LZ_L$ or $hh$) in the initial and/or final state, we have to multiply by hand an extra $1/\sqrt{2}$ onto the matrix element that we calculate in the usual way.\footnote{For two identical particles in the final state, this extra $1/\sqrt{2}$ accounts in advance for the fact that computing the total cross section involves integrating the squared matrix element over only \emph{half} of the solid angle.  For two identical particles in the initial state, this extra $1/\sqrt{2}$ accounts for the fact that properly (anti-)symmetrizing the initial-state wavefunction in the original quantum mechanical scattering formalism concentrates all of the initial flux into only the even (odd) partial waves, each of which thus contains \emph{twice} the incoming flux as for non-identical particle scattering.  (In any case the amplitudes that we work with here must also be invariant under time-reversal of the scattering process.)}

We already found:
\begin{equation}
	a_0(W^+_L W^-_L \to W^+_L W^-_L) \simeq - \frac{m_h^2}{8 \pi v^2}.
\end{equation}
The rest are:
\begin{align}
	a_0 (W^+_L W^-_L \leftrightarrow Z_L Z_L) 
		&\simeq - \frac{m_h^2}{8 \pi v^2} \cdot \frac{1}{\sqrt{8}} & 
		\qquad \qquad &(1/\sqrt{2} \ {\rm included}), \\
	a_0 (W^+_L W^-_L \leftrightarrow hh) 
		&\simeq - \frac{m_h^2}{8 \pi v^2} \cdot \frac{1}{\sqrt{8}} &
		\qquad \qquad &(1/\sqrt{2} \ {\rm included}), \\
	a_0 (Z_L Z_L \to Z_L Z_L) 
		&\simeq - \frac{m_h^2}{8 \pi v^2} \cdot \frac{3}{4} &
		\qquad \qquad &(1/\sqrt{2} \ {\rm included \ twice}), \\
	a_0 (Z_L Z_L \leftrightarrow hh) 
		&\simeq - \frac{m_h^2}{8 \pi v^2} \cdot \frac{1}{4} &
		\qquad \qquad &(1/\sqrt{2} \ {\rm included \ twice}), \\
	a_0 (hh \to hh) 
		&\simeq - \frac{m_h^2}{8 \pi v^2} \cdot \frac{3}{4} &
		\qquad \qquad &(1/\sqrt{2} \ {\rm included \ twice}),
\end{align}
where I'm everywhere ignoring pieces of order $g^2$ and pieces that fall with energy when $s \gg m_h^2$.  Assembling these into the \emph{coupled-channel matrix}\footnote{By construction, this matrix must be symmetric.} in the basis of initial or final states $(W_L^+ W_L^-, Z_L Z_L, hh)$ gives
\begin{equation}
	a_0 \simeq - \frac{m_h^2}{8 \pi v^2} \left( \begin{array}{ccc}
		1 & 1/\sqrt{8} & 1/\sqrt{8} \\
		1/\sqrt{8} & 3/4 & 1/4 \\
		1/\sqrt{8} & 1/4 & 3/4 \end{array} \right).
\end{equation}
The eigenvalues of the matrix in the parentheses are $3/2$, $1/2$, and $1/2$.  Selecting the largest of these (i.e., $3/2$)\footnote{The eigenvector corresponding to this largest eigenvalue is $(2 W^+_L W^-_L + Z_L Z_L + hh)/\sqrt{6}$.}, the perturbative unitarity bound $|{\rm Re} \, a_0| \leq 1/2$ becomes
\begin{equation}
	\left| - \frac{m_h^2}{8 \pi v^2} \cdot \frac{3}{2} \right| \leq 1/2 \qquad \Rightarrow \qquad
	\boxed{
		m_h^2 \leq \frac{8 \pi v^2}{3} \simeq (712~{\rm GeV})^2.
	}
\end{equation}

This is the upper bound on the Higgs boson mass that was famously first calculated by Lee, Quigg, and Thacker in 1977~\cite{Lee:1977yc,Lee:1977eg}.\footnote{Note that Lee, Quigg, and Thacker used $|a_0| \leq 1$, leading to an upper bound on $m_h$ larger by a factor of $\sqrt{2}$.  As mentioned previously, this is largely a matter of taste, and neither choice is more objectively ``correct'' than the other.  The choice to impose $|a_0| \leq 1$ in this situation is more conservative in the sense that it builds in more of a safety factor when deciding what maximum collider energy is needed to experimentally probe the underlying dynamics of electroweak symmetry breaking.}

\section{Application to the two-Higgs-doublet model}
\label{sec:2HDM}

Our first step away from the SM will be to consider longitudinal gauge boson scattering in the two-Higgs-doublet model (2HDM).\footnote{For an excellent and comprehensive review of the 2HDM, see Ref.~\cite{Branco:2011iw}.}  Everything in this section applies equally well to the Higgs sector of the Minimal Supersymmetric Standard Model.  The picture in the singlet extension of the SM is very similar and I'll comment on it briefly below.

\subsection{Some details of the SM Higgs-exchange diagrams}

To set up the calculation, let's first take a closer look at the SM Higgs exchange contributions to $W_L^+ W_L^-$ and $Z_L Z_L$ scattering.\footnote{For the remainder of these lectures, I'm going to ignore processes with Higgs bosons in the initial or final state.  In extended Higgs sectors, considering diagrams with initial- or final-state Higgs bosons becomes complicated for two reasons: first, simply because there are more Higgs bosons; and second, because the calculation of the four-point diagram involving four external Higgs bosons requires use of the four-Higgs vertices, which I want to avoid in these lectures for simplicity.  My purpose in including only the diagrams with external gauge bosons is to highlight that generic and powerful constraints on these models can be derived with a relatively small amount of calculational labour.}

For the channel $W^+_L W^-_L \to W^+_L W^-_L$, the Feynman diagrams involving SM Higgs exchange were shown already in Fig.~\ref{fig:WWstHiggs}.  The Higgs boson can show up in the $s$ channel and the $t$ channel, resulting in a contribution to the matrix element as given already in Eqs.~(\ref{eq:Mst1}--\ref{eq:Mhiggspart}), 
\begin{equation}
	\mathcal{M} = - \frac{1}{v^2} \left( \frac{s^2}{s - m_h^2} + \frac{t^2}{t - m_h^2} \right)
	\simeq - \frac{1}{v^2} \left( s + t + 2 m_h^2 + \cdots \right),
\end{equation}
where the approximation involves expanding the propagator denominators in a Taylor series for $s, t \gg m_h^2$ and dropping terms that fall with increasing collision energy.  The $(s + t)$ piece inside the parentheses is the piece that cancels the bad high-energy behaviour of the diagrams involving only gauge bosons.

\begin{figure}
\center
\includegraphics{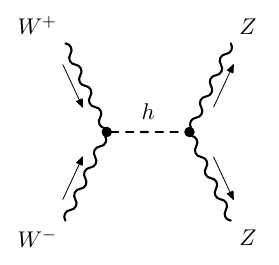} 
\caption{Diagram with an $s$-channel SM Higgs boson contributing to $W^+_L W^-_L \to Z_L Z_L$ scattering.}
\label{fig:WWZZhiggs}
\end{figure}

For the channel $W^+_L W^-_L \to Z_L Z_L$, there is only a single Feynman diagram involving SM Higgs exchange (in the $s$ channel) and it is shown in Fig.~\ref{fig:WWZZhiggs}.\footnote{Diagrams with $t$- or $u$-channel SM Higgs exchange do not appear for $W^+_L W^-_L \to Z_L Z_L$ because they would not conserve electric charge, as can be easily seen by attempting to draw them.}  The result from calculating this diagram (and multiplying by the extra factor of $1/\sqrt{2}$ required by the two identical $Z$ bosons in the final state) is,
\begin{equation}
	\mathcal{M} \cdot \frac{1}{\sqrt{2}} 
		= - \frac{1}{v^2} \left( \frac{s^2}{s - m_h^2} \right) \cdot \frac{1}{\sqrt{2}}
		\simeq - \frac{1}{v^2} \left( s + m_h^2 + \cdots \right) \cdot \frac{1}{\sqrt{2}}.
\end{equation}
The $s$ term inside the parentheses is the piece that cancels the bad high energy behaviour of the diagrams involving gauge bosons for this process.

\begin{figure}
\center
\includegraphics{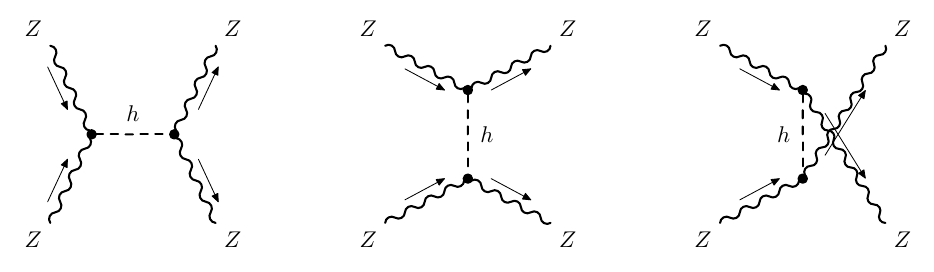}
\caption{Diagrams with $s$-, $t$-, and $u$-channel SM Higgs bosons contributing to $Z_L Z_L \to Z_L Z_L$ scattering.}
\label{fig:ZZhiggs}
\end{figure}

Finally for the channel $Z_L Z_L \to Z_L Z_L$, the three Feynman diagrams involving SM Higgs exchange are shown in Fig.~\ref{fig:ZZhiggs}.  The result from calculating these three diagrams (and multiplying by the two extra factors of $1/\sqrt{2}$ required by the two identical $Z$ bosons in both the initial state and the final state) is,
\begin{equation}
	\mathcal{M} \cdot \frac{1}{\sqrt{2}} \cdot \frac{1}{\sqrt{2}} 
		= - \frac{1}{v^2} \left( \frac{s^2}{s - m_h^2} + \frac{t^2}{t - m_h^2} + \frac{u^2}{u - m_h^2} 
			\right) \cdot \frac{1}{\sqrt{2}} \cdot \frac{1}{\sqrt{2}} 
		\simeq - \frac{1}{v^2} \left( s + t + u + 3 m_h^2 + \cdots \right) 
			\cdot \frac{1}{\sqrt{2}} \cdot \frac{1}{\sqrt{2}}.
\end{equation}
For this process, however, there are \emph{no pure-gauge contributions}, simply because the SM contains no (tree-level) $ZZZZ$, $ZZZ$, or $ZZ\gamma$ vertices.  This means that there is no bad high-energy behaviour for the Higgs-exchange diagrams to cancel, and indeed the Higgs-exchange diagrams do not grow with energy either, because\footnote{The more general version of this identity is $s + t + u = \sum_{i = 1}^4 m_i^2$, where $m_i$ are the masses of the four external particles.} $s + t + u = 4 M_Z^2$ (we are going to neglect the $4 M_Z^2$ piece for the same reason that we neglected the piece proportional to $M_W^2$ in the calculation of $WW$ scattering above).

With these results in hand, it is very straightforward to generalize the calculation to the case of the 2HDM.

\subsection{Field content and couplings in the 2HDM}
\label{sec:2HDMdefs}

The 2HDM contains two complex SU(2) doublets of scalar fields:
\begin{equation}
	\Phi_1 = \left( \begin{array}{c} \phi_1^+ \\ (v_1 + \phi_1^{0,r} + i \phi_1^{0,i})/\sqrt{2} \end{array} \right), \qquad 
	\Phi_2 = \left( \begin{array}{c} \phi_2^+ \\ (v_2 + \phi_2^{0,r} + i \phi_2^{0,i})/\sqrt{2} \end{array} \right).
\end{equation}
After electroweak symmetry breaking, the physical Higgs bosons are two CP-even neutral Higgs bosons $h^0$ and $H^0$ (with the usual convention that $h^0$ is the lighter of the two), one CP-odd neutral Higgs $A^0$, and a charged Higgs pair $H^{\pm}$.  Only the CP-even neutral Higgs bosons couple to vector boson pairs, and their interaction Feynman rules are
\begin{eqnarray}
	h^0 V_{\mu} V_{\nu}: && 2 i \frac{M_V^2}{v} \sin(\beta - \alpha) g_{\mu\nu}, \\
	H^0 V_{\mu} V_{\nu}: && 2 i \frac{M_V^2}{v} \cos(\beta - \alpha) g_{\mu\nu},
\end{eqnarray}
where $V = W$ or $Z$, $v_1^2 + v_2^2 = v^2 \simeq (246~{\rm GeV})^2$ is the SM Higgs vev as usual, $\beta$ is the mixing angle defined by $\tan\beta = v_2/v_1$, and $\alpha$ is the mixing angle that determines the composition of $h^0$ and $H^0$, i.e., $h^0 = -\sin\alpha \, \phi_1^{0,r} + \cos\alpha \, \phi_2^{0,r}$.  For comparison, the corresponding SM Higgs coupling $h V_{\mu} V_{\nu}$ is $2 i (M_V^2/v) g_{\mu\nu}$, i.e., the only change in going to the 2HDM is including the appropriate $\sin(\beta - \alpha)$ or $\cos(\beta-\alpha)$ factor.  This makes it convenient to define a \emph{coupling modification factor} $\kappa_{VV}^{\varphi}$, for the coupling of Higgs boson $\varphi$ to vector bosons $VV$, according to
\begin{equation}
	\varphi V_{\mu} V_{\nu}: \ 2 i \frac{M_V^2}{v} \kappa_{VV}^{\varphi} g_{\mu\nu}.
\end{equation}

It's now very straightforward to modify our SM calculation above to account for the presence of the two Higgs bosons that contribute to the $VV$ scattering diagrams.  This calculation will give us two interesting results: 
\begin{enumerate}
\item \emph{sum rules} for the Higgs boson couplings from the parts of the diagrams that go like $\mathcal{O}(E^2/v^2)$; and 
\item \emph{mass constraints} (which depend on the couplings) from the parts of the diagrams that do not grow with energy.
\end{enumerate}

\subsection{$W^+_L W^-_L \to W^+_L W^-_L$ calculation and results in the 2HDM}

\begin{figure}
\center
\includegraphics{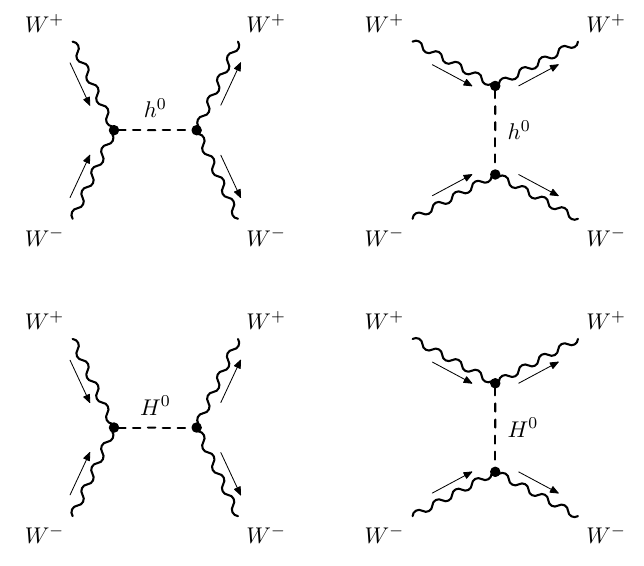}
\caption{Diagrams with $s$- and $t$-channel Higgs bosons in the 2HDM contributing to $W^+_L W^-_L \to W^+_L W^-_L$ scattering.}
\label{fig:WW2HDM}
\end{figure}

The Higgs-exchange diagrams that contribute to the process $W^+_L W^-_L \to W^+_L W^-_L$ in the 2HDM are shown in Fig.~\ref{fig:WW2HDM}; in particular, the SM diagrams just get duplicated to account for the fact that there are now two Higgs bosons, $h^0$ and $H^0$, that can contribute.  Indeed, the only differences compared to the SM calculation are the coupling modification factors $\kappa^{\varphi}_{WW}$ (one from each vertex, so two per diagram) and the two different Higgs masses that appear in the propagators.  The result is then easy to just read off from our SM calculation:  
\begin{eqnarray}
	\mathcal{M} &=& - \frac{1}{v^2} \left[ 
	(\kappa^{h^0}_{WW})^2 \left( \frac{s^2}{s - m_{h^0}^2} + \frac{t^2}{t - m_{h^0}^2} \right)
	+ (\kappa^{H^0}_{WW})^2 \left( \frac{s^2}{s - m_{H^0}^2} + \frac{t^2}{t - m_{H^0}^2} \right) \right] \\
	&\simeq& - \frac{1}{v^2} \left[ \left( (\kappa^{h^0}_{WW})^2 + (\kappa^{H^0}_{WW})^2 \right) (s + t)
	+ 2 (\kappa^{h^0}_{WW})^2 m_{h^0}^2 + 2 (\kappa^{H^0}_{WW})^2 m_{H^0}^2 + \cdots \right],
	\label{eq:2HDMWW}
\end{eqnarray}
where as usual in the second line we Taylor-expanded the propagator denominators in the high-scattering-energy limit and neglected terms that fall with increasing scattering energy.

From this expression we obtain a \emph{sum rule}: in order for the $\mathcal{O}(E^2/v^2)$ piece of these diagrams to successfully cancel the $\mathcal{O}(E^2/v^2)$ piece of the pure-gauge diagrams, it must be identical to the corresponding piece in the SM calculation; i.e., we must have
\begin{equation}
	\boxed{
	(\kappa^{h^0}_{WW})^2 + (\kappa^{H^0}_{WW})^2 = 1.
	}
\end{equation}
This is of course satisfied in the 2HDM because $\kappa^{h^0}_{WW} = \sin(\beta - \alpha)$ and $\kappa^{H^0}_{WW} = \cos(\beta - \alpha)$, and $\sin^2(\beta-\alpha) + \cos^2(\beta-\alpha) = 1$ by a trigonometry identity.  However, we see from the use of perturbative unitarity that this sum rule must hold more generally: in any model for which the diagrams in Fig.~\ref{fig:WW2HDM} capture all of the Higgs-boson contributions to $WW$ scattering, the couplings of the two contributing Higgs bosons must satisfy this relationship.  This is the case, for example, in the extension of the SM by a gauge-singlet scalar which mixes with the SM Higgs boson, creating two mass eigenstates.  It also generalizes to the case of more than two CP-even neutral Higgs bosons, in which the squares of the $\varphi WW$ coupling modification factors must always sum up to 1 if the Higgs-exchange diagrams are to successfully cancel the bad high-energy behaviour of the gauge-boson-only diagrams.\footnote{Models in which the Higgs degrees of freedom do not satisfy this requirement necessarily contain additional dynamics which should appear at or below the energy scale at which the (effective) theory violates perturbative unitarity.}

From Eq.~(\ref{eq:2HDMWW}) we also obtain a \emph{mass constraint}, as follows.  Combining the Higgs-exchange diagrams with the gauge-only diagrams and dropping terms that are suppressed by powers of $E$ in the high-energy limit (and contributions of order $g^2$, which are numerically small), the zeroth partial wave amplitude for longitudinal $WW$ scattering becomes
\begin{equation}
	a_0 = - \frac{1}{16 \pi v^2} \left[ 2 (\kappa^{h^0}_{WW})^2 m_{h^0}^2 
		+ 2 (\kappa^{H^0}_{WW})^2 m_{H^0}^2 \right].
\end{equation}
First, we take advantage of the sum rule to write 
\begin{equation}
	(\kappa^{h^0}_{WW})^2 = 1 - (\kappa^{H^0}_{WW})^2.
\end{equation}
Imposing $|{\rm Re} \, a_0| \leq 1/2$ and rearranging, we get:
\begin{equation}
	\boxed{
	(\kappa^{H^0}_{WW})^2 \leq \frac{4 \pi v^2 - m_{h^0}^2}{m_{H^0}^2 - m_{h^0}^2} 
	\simeq \left( \frac{870~{\rm GeV}}{m_{H^0}} \right)^2  \qquad {\rm or} \qquad
	m_{H^0} \lesssim \frac{870~{\rm GeV}}{| \kappa^{H^0}_{WW} |}.
	}
	\label{eq:2HDMmass}
\end{equation}
This result can be thought of as an upper bound on the $WW$ couplings of the heavier Higgs boson $H^0$ for a given $H^0$ mass, or as an upper bound on the $H^0$ mass for a fixed value of its coupling to $WW$ (which is itself intimately connected to the contribution of $H^0$ to electroweak symmetry breaking).

In other words, perturbative unitarity forces a \emph{decoupling behaviour}: the $H^0VV$ coupling is suppressed at high $H^0$ masses, and in the 2HDM we have
\begin{equation}
	\cos^2 (\beta - \alpha) \lesssim \frac{4 \pi v^2}{m_{H^0}^2} \qquad {\rm (at \ least)}.
\end{equation}
In fact, when one works out the 2HDM spectrum and couplings in detail, it turns out that the decoupling happens considerably faster~\cite{Gunion:2002zf}, with the $H^0VV$ coupling modification factor falling with increasing $H^0$ mass like
\begin{equation}
	\cos^2 (\beta - \alpha) \sim \frac{v^4}{m_{H^0}^4},
\end{equation}
but this cannot be seen just from the perturbative unitarity calculation.\footnote{In the SM plus a singlet, the decoupling behaviour just follows the perturbative unitarity bound in Eq.~(\ref{eq:2HDMmass}).  The practical upshot of this is that the heavy Higgs boson $H^0$ in the SM-plus-singlet model can generally have substantially larger couplings to vector boson pairs than that in the 2HDM.}

Calculating the amplitudes for $W^+_L W^-_L \to Z_L Z_L$ and $Z_L Z_L \to Z_L Z_L$ in the 2HDM, or adding additional Higgs doublets and/or singlets to the model, does not add any qualitatively new features to the discussion above.  Things get more interesting, however, in ``exotic'' extended Higgs sectors that contain SU(2) triplets or larger representations.  In the next two sections we will consider (1) the scalar septet model and (2) models with custodial symmetry in the scalar sector (the prototype for which is the Georgi-Machacek model).  Both of these preserve the electroweak rho parameter $\rho = 1$ at tree level even when the ``exotic'' scalar multiplet(s) have non-negligible vevs---which is necessary for the exotic scalars to have non-negligible $\varphi VV$ couplings since these couplings come from the $(\mathcal{D}_{\mu} \Phi)^\dagger \mathcal{D}^{\mu} \Phi$ terms in the Lagrangian.

\section{Application to the scalar septet model}
\label{sec:septet}

It has been known for a long time\footnote{See, e.g., footnote 1 of Ref.~\cite{Chanowitz:1985ug}.} that there are multiple representations of SU(2)$\times$U(1) that preserve $\rho \equiv M_W^2/M_Z^2 \cos^2 \theta_W = 1$ at tree level when they are used to break the electroweak symmetry.  The simplest of these is obviously the Higgs doublet with $T=1/2$ and $Y=1/2$,\footnote{I'm using the convention for the normalization of the hypercharge quantum number $Y$ that satisfies $Q = T^3 + Y$.  There is another convention in which $Q = T^3 + Y/2$.} as used in the SM; the next-smallest is a septet with $T=3$ and $Y=2$.\footnote{See homework question number 4 at the end of my TASI 2013 lectures~\cite{Logan:2014jla} for details.}  The scalar septet model was first implemented in a phenomenologically-viable way in Refs.~\cite{Hisano:2013sn,Kanemura:2013mc}, and most of the material in this section is based on Ref.~\cite{Harris:2017ecz}.

\subsection{Field content and couplings in the scalar septet model}

The scalar septet model contains a Higgs doublet $\Phi$ and a septet $X$, 
\begin{equation}
	\Phi = \left( \begin{array}{c} \phi^+ \\ \phi^0 \end{array} \right), \qquad \qquad
	X = \left( \begin{array}{c} \chi^{+5} \\ \chi^{+4} \\ \chi^{+3} \\ \chi^{+2} \\ \chi^{+1} \\ 
		\chi^0 \\ \chi^{-1} \end{array} \right),
\end{equation}
with vevs defined according to 
\begin{equation}
	\phi^0 = (v_{\phi} + \phi^{0,r} + i \phi^{0,i})/\sqrt{2}, \qquad \qquad
	\chi^0 = (v_{\chi} + \chi^{0,r} + i \chi^{0,i})/\sqrt{2}.
\end{equation}
The physical spectrum in the scalar sector is as follows:
\begin{itemize}
\item Two CP-even neutral scalars $h^0$ (lighter) and $H^0$ (heavier), from $\phi^{0,r}$ and $\chi^{0,r}$:
	\begin{eqnarray}
	h^0 &=& \cos\alpha \, \phi^{0,r} - \sin\alpha \, \chi^{0,r}, \\
	H^0 &=& \sin\alpha \, \phi^{0,r} + \cos\alpha \, \chi^{0,r};
	\end{eqnarray}
\item One CP-odd neutral scalar $A^0$ (and one neutral Goldstone boson $G^0$), from $\phi^{0,i}$ and $\chi^{0,i}$:
	\begin{eqnarray}
	G^0 &=& \cos\theta_7 \, \phi^{0,i} + \sin\theta_7 \, \chi^{0,i}, \\
	A^0 &=& -\sin\theta_7 \, \phi^{0,i} + \cos\theta_7 \, \chi^{0,i},
	\end{eqnarray}
where $\cos\theta_7 = v_{\phi}/v$ and $\sin\theta_7 = 4 v_{\chi}/v$, which are chosen this way because $M_W^2 = g^2 (v_{\phi}^2 + 16 v_{\chi}^2)/4$ in this model (i.e., $v_{\phi}^2 + 16 v_{\chi}^2 = v^2 \simeq (246~{\rm GeV})^2$);
\item Two singly-charged scalar pairs $H^{\pm}_f$ and $H^{\pm}_V$ (and one singly-charged Goldstone boson pair $G^{\pm}$), from $\phi^+$, $\chi^{+1}$, and $\chi^{-1}$ and their antiparticles:
	\begin{eqnarray}
	G^+ &=& \cos\theta_7 \, \phi^+ + \sin\theta_7 \left( \sqrt{\frac{5}{8}} \chi^{+1} 
		- \sqrt{\frac{3}{8}} \left(\chi^{-1}\right)^* \right), \\
	H^+_f &=& -\sin\theta_7 \, \phi^+ + \cos\theta_7 \left( \sqrt{\frac{5}{8}} \chi^{+1}
		- \sqrt{\frac{3}{8}} \left( \chi^{-1} \right)^* \right), \\
	H^+_V &=& \sqrt{\frac{3}{8}} \chi^{+1} + \sqrt{\frac{5}{8}} \left( \chi^{-1} \right)^*,
	\end{eqnarray}
where $H^+_f$ couples to fermion pairs but not vector boson pairs and $H^+_V$ couples to vector boson pairs but not fermion pairs. Note in general that there is nothing to prevent $H^+_f$ and $H^+_V$ from mixing to form mass eigenstates; we can define the lighter and heavier mass eigenstates respectively by
	\begin{eqnarray}
	H^+_1 &=& \cos\gamma \, H^+_f - \sin\gamma \, H^+_V, \\
	H^+_2 &=& \sin\gamma \, H^+_f + \cos\gamma \, H^+_V;
	\end{eqnarray}
\item One each of doubly, triply, quadruply, and pentuply charged scalar pairs:
	\begin{equation}
	H^{++} \equiv \chi^{+2}, \qquad \chi^{+3}, \qquad \chi^{+4}, \qquad \chi^{+5},
	\end{equation}
where $H^{++}$ couples to vector boson pairs but not fermion pairs, and the remaining states couple to neither vector boson pairs nor fermion pairs because of charge conservation.\footnote{These higher-charged states can decay through emission of $W$ bosons in cascades such as $\chi^{+5} \to \chi^{+4} W^+ \to \chi^{+3} W^+ W^+ \to H^{++} W^+ W^+ W^+$.}
\end{itemize}

For the purposes of vector boson scattering, we are interested in the scalars that have couplings $\varphi V_1 V_2$; the relevant scalars are $h^0$, $H^0$, $H^+_V$, and $H^{++}$.
The couplings of $h^0$ to vector boson pairs are,\footnote{Compare these to the SM Higgs couplings,
\begin{equation}
	h W_{\mu} W_{\nu}: \ \frac{2 i M_W^2}{v} g_{\mu\nu} \qquad \qquad
	h Z_{\mu} Z_{\nu}: \ \frac{2 i M_Z^2}{v} g_{\mu\nu}.
\end{equation}}
\begin{eqnarray}
	h^0 W^+_{\mu} W^-_{\nu}: && \frac{2 i M_W^2}{v} g_{\mu \nu}
		\left( \cos\theta_7 \cos\alpha - 4 \sin\theta_7 \sin\alpha \right)
		= \frac{2 i M_W^2}{v} g_{\mu \nu} \kappa^{h^0}_{WW}, \\
	h^0 Z_{\mu} Z_{\nu}: && \frac{2 i M_Z^2}{v} g_{\mu\nu}
		\left( \cos\theta_7 \cos\alpha - 4 \sin\theta_7 \sin\alpha \right)
		= \frac{2 i M_Z^2}{v} g_{\mu\nu} \kappa^{h^0}_{ZZ},
\end{eqnarray}
where the coupling modification factor is 
\begin{equation}
	\kappa^{h^0}_{WW} = \kappa^{h^0}_{ZZ} \equiv \kappa^{h^0}_{VV} = 
	\cos\theta_7 \cos\alpha - 4 \sin\theta_7 \sin\alpha.
\end{equation}
Similarly the couplings of $H^0$ to vector boson pairs are,
\begin{eqnarray}
	H^0 W^+_{\mu} W^-_{\nu}: && \frac{2 i M_W^2}{v} g_{\mu \nu}
		\left( \cos\theta_7 \sin\alpha + 4 \sin\theta_7 \cos\alpha \right)
		= \frac{2 i M_W^2}{v} g_{\mu \nu} \kappa^{H^0}_{WW}, \\
	H^0 Z_{\mu} Z_{\nu}: && \frac{2 i M_Z^2}{v} g_{\mu\nu}
		\left( \cos\theta_7 \sin\alpha + 4 \sin\theta_7 \cos\alpha \right)
		= \frac{2 i M_Z^2}{v} g_{\mu\nu} \kappa^{H^0}_{ZZ},
\end{eqnarray}
where the coupling modification factor is 
\begin{equation}
	\kappa^{H^0}_{WW} = \kappa^{H^0}_{ZZ} \equiv \kappa^{H^0}_{VV} = 
	\cos\theta_7 \sin\alpha + 4 \sin\theta_7 \cos\alpha.
\end{equation}
Notice that, in contrast to the case in the 2HDM, $(\kappa^{h^0}_{VV})^2 + (\kappa^{H^0}_{VV})^2 \neq 1$!  This is going to be important in a minute.

The coupling of $H^+_V$ to vector boson pairs is
\begin{equation}
	H^+_V W^-_{\mu} Z_{\nu}: \ \frac{2 i M_W M_Z}{v} g_{\mu\nu} 
	\left( -\sqrt{15} \sin\theta_7 \right),
\end{equation}
and the coupling of $H^{++}$ to vector boson pairs is
\begin{equation}
	H^{++} W^-_{\mu} W^-_{\nu}: \ \frac{2 i M_W^2}{v} g_{\mu\nu}
	\left( \sqrt{15} \sin\theta_7 \right).
\end{equation}
Even though there is (obviously) no SM Higgs coupling to $W^-Z$ or to $W^-W^-$, we can define coupling modification factors in analogy to those of the CP-even neutral Higgses by writing\footnote{If we allow mixing of $H^+_V$ and $H^+_f$, the coupling modification factors of the two singly-charged mass eigenstates become
\begin{equation}
	\kappa^{H^+_1}_{WZ} = \sqrt{15} \sin\theta_7 \sin\gamma, \qquad \qquad
	\kappa^{H^+_2}_{WZ} = -\sqrt{15} \sin\theta_7 \cos\gamma.
\end{equation}}
\begin{equation}
	\varphi V_{1 \mu} V_{2 \nu}: \ \frac{2 i M_{V_1} M_{V_2}}{v} g_{\mu\nu} 
		\kappa^{\varphi}_{V_1V_2}.
\end{equation}
This lets us define\footnote{The fact that these two coupling modification factors are the same up to a minus sign has no special significance outside of the context of this particular model.}
\begin{equation}
	\kappa^{H^+_V}_{WZ} = -\sqrt{15} \sin\theta_7, \qquad \qquad
	\kappa^{H^{++}}_{WW} = \sqrt{15} \sin\theta_7.
\end{equation}

In the 2HDM we had $(\kappa^{h^0}_{VV})^2 + (\kappa^{H^0}_{VV})^2 = 1$; in fact this sum rule was \emph{required} by perturbative unitarity of longitudinal vector boson scattering in order for the Higgs-exchange diagrams to cancel the bad high energy behaviour of the gauge-boson-only diagrams.  Here we have (using $c$ for cosine and $s$ for sine),
\begin{eqnarray}
	\left( \kappa^{h^0}_{VV} \right)^2 + \left( \kappa^{H^0}_{VV} \right)^2
	&=& (c_7 c_{\alpha} - 4 s_7 s_{\alpha} )^2 + (c_7 s_{\alpha} + 4 s_7 c_{\alpha})^2 \nonumber \\
	&=&c_7^2 c_{\alpha}^2 - 8 c_7 s_7 c_{\alpha} s_{\alpha} + 16 s_7^2 s_{\alpha}^2 \nonumber \\
	&& \! \! \! + \, c_7^2 s_{\alpha}^2 + 8 c_7 s_7 c_{\alpha} s_{\alpha} + 16 s_7^2 c_{\alpha}^2 \nonumber \\
	&=& c_7^2 + 16 s_7^2 \geq 1, \qquad \qquad {\rm independent \ of \ \alpha}.
\end{eqnarray}
How is this reconciled with perturbative unitarity of longitudinal vector boson scattering?  The answer is that in the scalar septet model there are \emph{additional scalars} that participate in the vector boson scattering: $H^+_V$ and $H^{++}$.  Recall that in the 2HDM, $H^+$ couples to fermion pairs but does not couple to vector boson pairs at tree level, so this feature is genuinely new, as (obviously) is the presence of the doubly-charged Higgs boson.

\subsection{Sum rules for couplings in the scalar septet model}

\begin{figure}
\center
\includegraphics{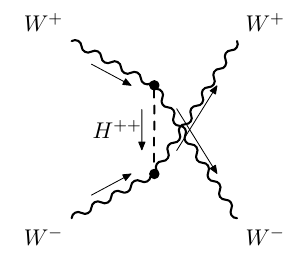} 
\caption{Diagram involving a $u$-channel $H^{++}$ that contributes to $W^+_L W^-_L \to W^+_L W^-_L$ in the scalar septet model.}
\label{fig:WWH2p}
\end{figure}

To see how this plays out, we go ahead and calculate the matrix element for the diagrams involving Higgs exchange.  First consider $W^+_L W^-_L \to W^+_L W^-_L$.  There are four diagrams involving $s$- and $t$-channel exchange of $h^0$ and $H^0$, which look exactly the same as in the 2HDM and were shown already in Fig.~\ref{fig:WW2HDM}.  There is also a new diagram involving $u$-channel exchange of $H^{++}$, as shown in Fig.~\ref{fig:WWH2p}.

Calculating these diagrams gives the Higgs-exchange piece of the matrix element,
\begin{eqnarray}
	\mathcal{M} &=& - \frac{1}{v^2} \left[ 
	\left( \kappa^{h^0}_{WW} \right)^2 \left( \frac{s^2}{s - m_{h^0}^2} + \frac{t^2}{t - m_{h^0}^2} \right)
	+ \left( \kappa^{H^0}_{WW} \right)^2 \left( \frac{s^2}{s - m_{H^0}^2} 
		+ \frac{t^2}{t - m_{H^0}^2} \right) \right. \nonumber \\
	&& \left. \qquad \ \ 
	+ \left( \kappa^{H^{++}}_{WW} \right)^2 \left( \frac{u^2}{u - m_{H^{++}}^2} \right) \right].
\end{eqnarray}
Expanding this in the high-energy limit gives,
\begin{eqnarray}
	\mathcal{M} &\simeq& - \frac{1}{v^2} \left[
	\left( \left( \kappa^{h^0}_{WW} \right)^2 + \left( \kappa^{H^0}_{WW} \right)^2 \right) (s + t)
	+ \left( \kappa^{H^{++}}_{WW} \right)^2 u \right. \nonumber  \\
	&& \left. \qquad \ \ 
	+ \, 2 \left( \kappa^{h^0}_{WW} \right)^2 m_{h^0}^2 
	+ 2 \left( \kappa^{H^0}_{WW} \right)^2 m_{H^0}^2
	+ \left( \kappa^{H^{++}}_{WW} \right)^2 m_{H^{++}}^2 \right].
	\label{eq:WWWWseptet}
\end{eqnarray}

We need the first line to equal $- \frac{1}{v^2} \left[ s + t \right]$ in order to cancel the $\mathcal{O}(E^2/v^2)$ part of the gauge-boson-only diagrams---this requirement is going to give us a more general version of the sum rule that we found in the 2HDM.  To see how this goes, we first use $s + t + u = 4 M_W^2 \simeq 0$ in our approximation of neglecting terms of order $M_W^2/v^2 \sim g^2$.  Then we get $u \simeq - (s + t)$, which allows us to combine the term coming from $H^{++}$ exchange with the terms coming from $h^0$ and $H^0$.  Unitarity thus imposes the \emph{sum rule} in the scalar septet model,
\begin{equation}
	\boxed{
	\left( \kappa^{h^0}_{WW} \right)^2 + \left( \kappa^{H^0}_{WW} \right)^2 
	- \left( \kappa^{H^{++}}_{WW} \right)^2 = 1.
	}
\end{equation}
The minus sign in this equation came from $u \simeq - (s+t)$, thereby letting us factor out an overall $(s + t)$ and giving us a chance to reconcile the fact that the sum of the squares of the $h^0$ and $H^0$ coupling modification factors is generally greater than one in this model.  In terms of the actual values of the coupling modification factors in the scalar septet model, we have
\begin{equation}
	c_7^2 + 16 s_7^2 - \left( \sqrt{15} s_7 \right)^2 = c_7^2 + s_7^2 = 1. \qquad \qquad {\rm Success!}
\end{equation}

\begin{figure}
\center
\includegraphics{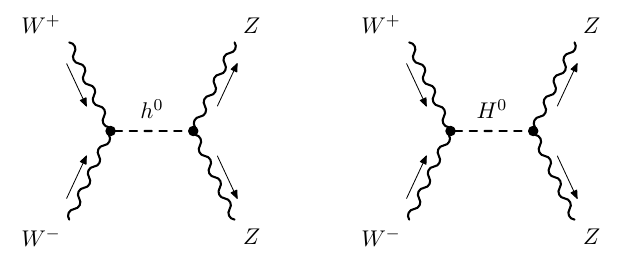} 
\caption{Diagrams for $s$-channel exchange of $h^0$ and $H^0$ in $W^+_L W^-_L \to Z_L Z_L$ scattering in the scalar septet model.}
\label{fig:WWZZhH}
\end{figure}

\begin{figure}
\center
\includegraphics{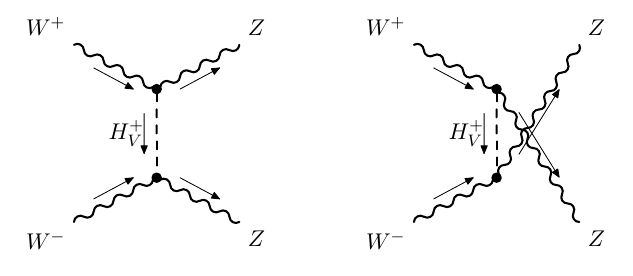} 
\caption{Diagrams for $t$- and $u$-channel exchange of $H^+_V$ in $W^+_L W^-_L \to Z_L Z_L$ scattering in the scalar septet model.}
\label{fig:WWZZHptu}
\end{figure}

The doubly-charged scalar does not solve all our problems by itself.  Consider now $W^+_L W^-_L \to Z_L Z_L$.  As in the SM, charge conservation requires that $h^0$ and $H^0$ contribute to this process only in the $s$ channel, as shown in Fig.~\ref{fig:WWZZhH}.  But now there are also two new diagrams involving $t$- and $u$-channel exchange of $H^+_V$, as shown in Fig.~\ref{fig:WWZZHptu}.\footnote{I am ignoring the possible $H^+_V$--$H^+_f$ mixing here; it would be straightforward to add it back in.}

Calculating the matrix elements, using the fact that $\kappa^{h^0,H^0}_{WW} = \kappa^{h^0,H^0}_{ZZ} \equiv \kappa^{h^0,H^0}_{VV}$, and multiplying on the factor of $1/\sqrt{2}$ to account for the two identical $Z$ bosons in the final state, we get 
\begin{eqnarray}
	\mathcal{M} \cdot \frac{1}{\sqrt{2}} &=& - \frac{1}{v^2} \cdot \frac{1}{\sqrt{2}} \left[ 
	\left( \kappa^{h^0}_{VV} \right)^2 \frac{s^2}{s - m_{h^0}^2}
	+ \left( \kappa^{H^0}_{VV} \right)^2 \frac{s^2}{s - m_{H^0}^2}
	+ \left( \kappa^{H^+_V}_{WZ} \right)^2 \left( \frac{t^2}{t - m_{H^+_V}^2} 
		+ \frac{u^2}{u - m_{H^+_V}^2} \right) \right]  \\
	&\simeq& - \frac{1}{v^2} \cdot \frac{1}{\sqrt{2}} \left[ 
		\left( \left( \kappa^{h^0}_{VV} \right)^2 + \left( \kappa^{H^0}_{VV} \right)^2 \right) s
		+ \left( \kappa^{H^+_V}_{WZ} \right)^2 \left( t + u \right) \right. \nonumber \\
	&& \left. \qquad \qquad \quad
		+ \left( \kappa^{h^0}_{VV} \right)^2 m_{h^0}^2 
		+ \left( \kappa^{H^0}_{VV} \right)^2 m_{H^0}^2
		+ 2 \left( \kappa^{H^+_V}_{WZ} \right)^2 m_{H^+_V}^2 \right].
\end{eqnarray}
We need the first line to equal $-\frac{1}{v^2} \frac{1}{\sqrt{2}} s$ in order to cancel the $\mathcal{O}(E^2/v^2)$ part of the gauge-boson-only diagrams---this requirement is going to give us another more general version of the sum rule that we found in the 2HDM.
Again we can use $s + t + u \simeq 0$ to combine the terms of $\mathcal{O}(E^2/v^2)$ by writing $(t + u) \simeq -s$ and combining the term coming from $H^+_V$ exchange with the other term.  Unitarity then imposes a \emph{second} sum rule in the scalar septet model,\footnote{It should be evident to you how to generalize this sum rule to the (hypothetical) situation in which $\kappa^{\varphi^0}_{WW} \neq \kappa^{\varphi^0}_{ZZ}$ just by counting the coupling factors in the diagrams of Fig.~\ref{fig:WWZZhH}.}
\begin{equation}
	\boxed{
	\left( \kappa^{h^0}_{VV} \right)^2 + \left( \kappa^{H^0}_{VV} \right)^2 
	- \left( \kappa^{H^+_V}_{WZ} \right)^2 = 1.
	}
\end{equation}
Once again, the minus sign in this equation came from $(t + u) \simeq -s$, thereby letting us factor out an overall $s$ and compensating for the fact that the sum of the squares of the $h^0$ and $H^0$ coupling modification factors is generally greater than one in this model.  In terms of the actual values of the coupling modification factors in the scalar septet model, we have
\begin{equation}
	c_7^2 + 16 s_7^2 - \left( -\sqrt{15} s_7 \right)^2 = c_7^2 + s_7^2 = 1.
\end{equation}
This looks like a nice little miracle, but remember that it is required to hold in any renormalizable model, so it should not really come as a surprise that it works.

As mentioned before, the gauge-boson-only contribution to $Z_L Z_L \to Z_L Z_L$ scattering is in fact zero, so we don't get to write down a third sum rule from considering this channel.

\subsection{Mass and coupling upper bounds in the scalar septet model}

We can use the pieces of the matrix elements that do not grow with energy to derive bounds on combinations of the scalar masses and couplings.  From $W^+_L W^-_L \to W^+_L W^-_L$ scattering, setting the second line of Eq.~(\ref{eq:WWWWseptet}) equal to $16 \pi a_0$ and taking $|{\rm Re} \, a_0| \leq 1/2$ yields
\begin{equation}
	2 \left( \kappa^{h^0}_{WW} \right)^2 m_{h^0}^2 
	+ 2 \left( \kappa^{H^0}_{WW} \right)^2 m_{H^0}^2
	+ \left( \kappa^{H^{++}}_{WW} \right)^2 m_{H^{++}}^2 
	\leq 8 \pi v^2.
	\label{eq:massboundseptet}
\end{equation}
We want to simplify this to get, e.g., an upper bound on $s_7 \equiv 4 v_{\chi}/v$ (where remember that $v_{\chi}$ is the septet vev) as a function of some useful scalar mass (we will choose $m_{H^{++}}$).  We therefore choose as our target an upper bound on $\kappa^{H^{++}}_{WW}$, which is proportional to $s_7$.  We already know from our first sum rule that 
\begin{equation}
	\left( \kappa^{h^0}_{WW} \right)^2 + \left( \kappa^{H^0}_{WW} \right)^2 
		= 1 + \left( \kappa^{H^{++}}_{WW} \right)^2.
\end{equation}
For a \emph{conservative} upper bound on $\kappa^{H^{++}}_{WW}$ (i.e., a simplified bound that must always be satisfied, even if the full complicated version may be even stronger), let's choose the mixing angle $\alpha$ to give $\kappa^{H^{++}}_{WW}$ as much wiggle room as possible: for $m_{H^0} > m_{h^0}$, this amounts to dialing $\kappa^{H^0}_{WW}$ to zero.  Then we can write $\left( \kappa^{h^0}_{WW} \right)^2 = 1 - \left( \kappa^{H^{++}}_{WW} \right)^2$.  Plugging this into Eq.~(\ref{eq:massboundseptet}) with $\kappa^{H^0}_{WW} = 0$ gives
\begin{equation}
	2 \left[ 1 - \left( \kappa^{H^{++}}_{WW} \right)^2 \right] m_{h^0}^2 
	+ \left( \kappa^{H^{++}}_{WW} \right)^2 m_{H^{++}}^2 \leq 8 \pi v^2,
\end{equation}
which can be rearranged to get
\begin{equation}
	\left( \kappa^{H^{++}}_{WW} \right)^2 
	\leq \frac{8 \pi v^2 - 2 m_{h^0}^2}{m_{H^{++}}^2 + 2 m_{h^0}^2}
	\simeq \frac{8 \pi v^2}{m_{H^{++}}^2},
\end{equation}
where we're assuming that $m_{H^{++}} \gg m_{h^0} \simeq 125$~GeV.  Using $\kappa^{H^{++}}_{WW} = \sqrt{15} \sin\theta_7$, we get the upper bound,
\begin{equation}
	s_7^2 \lesssim \frac{8 \pi v^2}{15 m_{H^{++}}^2} 
	\simeq \left( \frac{315~{\rm GeV}}{m_{H^{++}}} \right)^2.
\end{equation}

As in the case of the 2HDM, we see a \emph{decoupling behaviour}: the ``exotic'' septet vev, which directly controls the coupling of $H^{++}$ (and $H^+_V$) to gauge boson pairs, must fall with increasing $H^{++}$ mass at least as fast as
\begin{equation}
	\sin^2\theta_7 \sim \frac{ {\rm (number)} \cdot v^2}{m_{H^{++}}^2}.
\end{equation}
This can alternatively be interpreted as an upper bound on $m_{H^{++}}$ for a given value of $s_7$; for example, if experiments were to measure deviations in the couplings of the SM-like Higgs boson $h^0$ that indicated a nonzero value for $s_7$, the size of these deviations would set an upper bound on the mass of the doubly-charged scalar in this model (which is always good news if your collider has limited energy reach).

Recall that I said in the previous section that the 2HDM actually decouples faster (i.e., with a higher power of $v^2/m^2$) than the bound deduced from perturbative unitarity.  I don't actually know whether the scalar septet model decouples with a higher power of $v^2/m^2$ than required by the perturbative unitarity bound because I have not worked through that aspect of the model in detail.\footnote{Such a calculation is made somewhat more complicated by the fact that the phenomenologically-viable implementations of the scalar septet model in Refs.~\cite{Hisano:2013sn,Kanemura:2013mc} both require the presence of additional new physics (either through a higher-dimension operator coming from extra scalar content or through an extension of the gauge sector) to get rid of a stray Goldstone boson.}

\section{Application to the Georgi-Machacek model}
\label{sec:GM}

The Georgi-Machacek (GM) model~\cite{Georgi:1985nv,Chanowitz:1985ug} is the simplest nontrivial implementation of an alternative approach to maintain $\rho = 1$ at tree level by imposing an additional SU(2) global symmetry on the Higgs sector so that the so-called custodial symmetry is preserved after electroweak symmetry breaking.  Perturbative unitarity in this class of models was studied in detail in Ref.~\cite{Falkowski:2012vh}; this section also draws on material in Ref.~\cite{Logan:2015xpa}.

\subsection{Field content and couplings in the GM model}

The GM model contains a Higgs doublet $\Phi$, a complex triplet $X$, and a real triplet $\Xi$,
\begin{equation}
	\Phi = \left( \begin{array}{c} \phi^+ \\ \phi^0 \end{array} \right), \qquad \qquad
	X = \left( \begin{array}{c} \chi^{++} \\ \chi^+ \\ \chi^0 \end{array} \right), \qquad \qquad
	\Xi = \left( \begin{array}{c} \xi^+ \\ \xi^0 \\ \xi^- \end{array} \right),
\end{equation}
with their hypercharge assignments chosen strategically to obtain the electric charges as shown.

An extra global SU(2)$_R$ symmetry is imposed by hand on the scalar potential; this is made manifest by writing the doublet (under the gauged SU(2)$_L$) as a bi-doublet and combining the triplets into a bi-triplet under the larger SU(2)$_L \times$SU(2)$_R$ symmetry,
\begin{equation}
	\Phi = \left( \begin{array}{cc} \phi^{0*} & \phi^+ \\
						-\phi^{+*} & \phi^0 \end{array} \right), \qquad \qquad
	X = \left( \begin{array}{ccc} \chi^{0*} & \xi^+ & \chi^{++} \\
						-\chi^{+*} & \xi^0 & \chi^+ \\
						\chi^{++*} & -\xi^{+*} & \chi^0 \end{array} \right),
\end{equation}
and constructing the scalar potential out of these.  In this notation the vevs are
\begin{equation}
	\langle \Phi \rangle = \left( \begin{array}{cc} v_{\phi}/\sqrt{2} & 0 \\
									0 & v_{\phi}/\sqrt{2} \end{array} \right), \qquad \qquad
	\langle X \rangle = \left( \begin{array}{ccc} v_{\chi} & 0 & 0 \\
									0 & v_{\chi} & 0 \\
									0 & 0 & v_{\chi} \end{array} \right),
\end{equation}
preserving the \emph{diagonal subgroup} SU(2)$_c$ (the custodial symmetry) of SU(2)$_L \times$SU(2)$_R$.  Note that this implies that $\langle \xi^0 \rangle = \langle \chi^0 \rangle$: this is \emph{not} totally general and is a reflection of the preserved custodial symmetry.  The equality of the vevs of the two triplets is what ensures that $\rho = 1$ at tree level.\footnote{For the details behind this statement, see Sec.~5 of my TASI 2013 lectures~\cite{Logan:2014jla}.}  

The physical spectrum in the scalar sector is as follows:
\begin{itemize}
\item Three CP-even neutral scalars $h^0$, $H^0$, and $H_5^0$, from $\phi^{0,r}$, $\chi^{0,r}$, and $\xi^{0,r}$:\footnote{As usual we define $\phi^0 = (v_{\phi} + \phi^{0,r} + i \phi^{0,i})/\sqrt{2}$; in this model we also have $\chi^0 = v_{\chi} + (\chi^{0,r} + i \chi^{0,i})/\sqrt{2}$ and $\xi^0 = v_{\chi} + \xi^{0,r}$.}
	\begin{eqnarray}
	h^0 &=& \cos\alpha \, \phi^{0,r} - \sin\alpha \left( \frac{1}{\sqrt{3}} \xi^{0,r} 
			+ \sqrt{\frac{2}{3}} \chi^{0,r} \right), \\
	H^0 &=& \sin\alpha \, \phi^{0,r} + \cos\alpha \left( \frac{1}{\sqrt{3}} \xi^{0,r} 
			+ \sqrt{\frac{2}{3}} \chi^{0,r} \right), \\
	H_5^0 &=& \sqrt{\frac{2}{3}} \xi^{0,r} - \frac{1}{\sqrt{3}} \chi^{0,r};
	\end{eqnarray}
\item One CP-odd neutral scalar $H_3^0$ (and one neutral Goldstone boson $G^0$) from $\phi^{0,i}$ and $\chi^{0,i}$:
	\begin{eqnarray}
		G^0 &=& \cos\theta_H \, \phi^{0,i} + \sin\theta_H \, \chi^{0,i}, \\
		H_3^0 &=& -\sin\theta_H \, \phi^{0,i} + \cos\theta_H \, \chi^{0,i},
	\end{eqnarray}
where $\cos\theta_H = v_{\phi}/v$ and $\sin\theta_H = \sqrt{8} v_{\chi}/v$, which are chosen this way because $M_W^2 = g^2 ( v_{\phi}^2 + 8 v_{\chi}^2 )/4$ in this model (i.e., $v_{\phi}^2 + 8 v_{\chi}^2 = v^2 \simeq (246~{\rm GeV})^2$);
\item Two singly-charged scalar pairs $H_3^{\pm}$ and $H_5^{\pm}$ (and one singly-charged Goldstone boson pair $G^{\pm}$), from $\phi^{\pm}$, $\chi^{\pm}$, and $\xi^{\pm}$:
	\begin{eqnarray}
		G^+ &=& \cos\theta_H \, \phi^+ + \sin\theta_H \left( \frac{\chi^+ + \xi^+}{\sqrt{2}} \right), \\
		H_3^+ &=& -\sin\theta_H \, \phi^+ + \cos\theta_H \left( \frac{\chi^+ + \xi^+}{\sqrt{2}} \right), \\
		H_5^+ &=& \frac{\chi^+ - \xi^+}{\sqrt{2}}, 
	\end{eqnarray}
where $H_3^+$ couples to fermion pairs but not vector boson pairs and $H_5^+$ couples to vector boson pairs but not fermion pairs; and
\item One doubly-charged scalar pair:
	\begin{equation}
		H_5^{++} = \chi^{++}.
	\end{equation}
\end{itemize}

These states can be grouped by their transformation properties under the custodial SU(2)$_c$, as follows.  The (bi-)doublet transforms as a $2 \otimes 2$ representation under SU(2)$_L \times$SU(2)$_R$, which breaks down to $3 \oplus 1$ under the diagonal subgroup SU(2)$_c$.  Similarly, the bi-triplet transforms as a $3 \otimes 3$ under SU(2)$_L \times$SU(2)$_R$, and breaks down to $5 \oplus 3 \oplus 1$ under SU(2)$_c$.  The two custodial singlets (particles transforming in the $1$ representation of SU(2)$_c$) mix to form $h^0$ and $H^0$.  The two custodial triplets (particles transforming in the $3$ representation of SU(2)$_c$) mix to form the Goldstone bosons $(G^+, G^0, G^-)$ along with a \emph{custodial triplet} of physical states, $(H_3^+, H_3^0, H_3^-)$.\footnote{Something very similar happens in the 2HDM.  In that model, the two SU(2)$_L$ doublets yield two custodial singlets and two custodial triplets, which respectively form the CP-even neutral Higgs bosons $h^0$ and $H^0$ and the Goldstone bosons and physical states $(H^+, A^0, H^-)$.  In the 2HDM, however, the custodial symmetry is usually explicitly broken by terms in the scalar potential, with the result that $H^{\pm}$ and $A^0$ are not degenerate in mass.}  The single \emph{custodial fiveplet} (particles transforming in the $5$ representation of SU(2)$_c$) contains the physical states $(H_5^{++}, H_5^+, H_5^0, H_5^-, H_5^{--})$.

The most important feature of the GM model is that custodial symmetry is preserved in the scalar sector.  This custodial symmetry prevents mixing between states that transform in different representations of SU(2)$_c$; for example, $H_3^+$ (which couples to fermion pairs but not to vector boson pairs) does \emph{not} mix with $H_5^+$ (which couples to vector boson pairs but not to fermion pairs).\footnote{Custodial symmetry in the GM model is violated at one loop, which induces mixing between states in different custodial multiplets, but this mixing is constrained to be very small by other experimental constraints on custodial symmetry violation~\cite{Keeshan:2018ypw}.}  This is in contrast to the scalar septet model in which $H^+_f$ and $H^+_V$ are allowed to mix in general.  Furthermore, $H_3^{\pm}$ and $H_3^0$ both have the same mass $m_3$, and $H_5^{\pm\pm}$, $H_5^{\pm}$, and $H_5^0$ all have the same mass $m_5$.

For our calculation of the Higgs contributions to longitudinal vector boson scattering, we will be interested in the $\varphi V_1 V_2$ couplings.  The relevant scalars that couple to vector boson pairs are $h^0$, $H^0$, $H_5^0$, $H_5^+$, and $H_5^{++}$.\footnote{Compare the scalar septet model, in which the relevant scalars were $h^0$, $H^0$, $H_V^+$, and $H^{++}$.}  The couplings are:\footnote{Compare to the scalar septet model, in which $\kappa^{H^+_V}_{WZ} = -\sqrt{15} \sin\theta_7$ and $\kappa^{H^{++}}_{WW} = \sqrt{15} \sin\theta_7$ and there is no analogue of $H_5^0$: evidently the cancellation of the $\mathcal{O}(E^2/v^2)$ pieces of the amplitudes is going to work a little bit differently in the GM model.}
\begin{eqnarray}
	\left. \begin{array}{r l} 
		h^0 W^+_{\mu} W^-_{\nu}: 
			& \frac{2 i M_W^2}{v} g_{\mu\nu} \kappa^{h^0}_{WW} \\
		h^0 Z_{\mu} Z_{\nu}: 
			& \frac{2 i M_Z^2}{v} g_{\mu\nu} \kappa^{h^0}_{ZZ} 
		\end{array} \right\}
	&& \kappa^{h^0}_{WW} = \kappa^{h^0}_{ZZ} \equiv \kappa^{h^0}_{VV} 
		= \cos\alpha \cos\theta_H - \sqrt{\frac{8}{3}} \sin\alpha \sin\theta_H, \\
	\left. \begin{array}{r l} 
		H^0 W^+_{\mu} W^-_{\nu}: 
			& \frac{2 i M_W^2}{v} g_{\mu\nu} \kappa^{H^0}_{WW} \\
		H^0 Z_{\mu} Z_{\nu}:
			& \frac{2 i M_Z^2}{v} g_{\mu\nu} \kappa^{H^0}_{ZZ}
		\end{array} \right\}
	&& \kappa^{H^0}_{WW} = \kappa^{H^0}_{ZZ} \equiv \kappa^{H^0}_{VV}
		= \sin\alpha \cos\theta_H + \sqrt{\frac{8}{3}} \cos\alpha \sin\theta_H, \\
	\left. \begin{array}{r l}
		H_5^0 W^+_{\mu} W^-_{\mu}:
			& \frac{2 i M_W^2}{v} g_{\mu\nu} \kappa^{H_5^0}_{WW} \\
		H_5^0 Z_{\mu} Z_{\nu}:
			& \frac{2 i M_Z^2}{v} g_{\mu\nu} \kappa^{H_5^0}_{ZZ}
		\end{array} \right\}
	&& \begin{array}{l}
		\kappa^{H_5^0}_{WW} = \frac{1}{\sqrt{3}} \sin\theta_H, \\
		\kappa^{H_5^0}_{ZZ} = - \frac{2}{\sqrt{3}} \sin\theta_H,
		\end{array} \qquad {\rm different!} \\
	\left. H_5^+ W^-_{\mu} Z_{\nu}: \ 
		\frac{2 i M_W M_Z}{v} g_{\mu\nu} \kappa^{H_5^+}_{WZ} \right\}
	&& \kappa^{H_5^+}_{WZ} = - \sin\theta_H, \\
	\left. H_5^{++} W^-_{\mu} W^-_{\nu}: \  
		\frac{2 i M_W^2}{v} g_{\mu\nu} \kappa^{H_5^{++}}_{WW} \right\}
	&& \kappa^{H_5^{++}}_{WW} = \sqrt{2} \sin\theta_H.
\end{eqnarray}

As in the scalar septet model, we do \emph{not} have $(\kappa^{h^0}_{VV})^2 + (\kappa^{H^0}_{VV})^2 = 1$:
\begin{eqnarray}
	\left( \kappa^{h^0}_{VV} \right)^2 + \left( \kappa^{H^0}_{VV} \right)^2 
	&=& \left( c_{\alpha} c_H - \sqrt{\frac{8}{3}} s_{\alpha} s_H \right)^2
		+ \left( s_{\alpha} c_H + \sqrt{\frac{8}{3}} c_{\alpha} s_H \right)^2 \nonumber \\
	&=& c_{\alpha}^2 c_H^2 - 2 \cdot \sqrt{\frac{8}{3}} c_{\alpha} s_{\alpha} c_H s_H 
		+ \frac{8}{3} s_{\alpha}^2 s_H^2 \nonumber \\
	&& \!\!\!\! + \, s_{\alpha}^2 c_H^2 + 2 \cdot \sqrt{\frac{8}{3}} c_{\alpha} s_{\alpha} c_H s_H 
		+ \frac{8}{3} c_{\alpha}^2 s_H^2 \nonumber \\
	&=& c_H^2 + \frac{8}{3} s_H^2 \geq 1,  \qquad \qquad {\rm independent \ of \ \alpha}.
\end{eqnarray}

\subsection{Sum rules and mass and coupling upper bounds in the GM model}

Let's see how the unitarity cancellations play out.  We'll start with $W^+_L W^-_L \to W^+_L W^-_L$.  There are $s$- and $t$-channel diagrams involving $h^0$, $H^0$, and now also $H_5^0$, along with a $u$-channel diagram involving $H_5^{++}$.  The matrix element is,
\begin{eqnarray}
	\mathcal{M} &=& - \frac{1}{v^2} \left[ 
		\left( \kappa^{h^0}_{WW} \right)^2 
			\left( \frac{s^2}{s - m_{h^0}^2} + \frac{t^2}{t - m_{h^0}^2} \right)
		+ \left( \kappa^{H^0}_{WW} \right)^2
			\left( \frac{s^2}{s - m_{H^0}^2} + \frac{t^2}{t - m_{H^0}^2} \right)
		\right. \nonumber \\
		&& \left. \qquad \ \ 
		+ \left( \kappa^{H_5^0}_{WW} \right)^2
			\left( \frac{s^2}{s - m_5^2} + \frac{t^2}{t - m_5^2} \right)
		+ \left( \kappa^{H_5^{++}}_{WW} \right)^2
			\left( \frac{u^2}{u - m_5^2} \right) \right] \\
		&\simeq& - \frac{1}{v^2} \left[
			\left( \left( \kappa^{h^0}_{WW} \right)^2 + \left( \kappa^{H^0}_{WW} \right)^2
				+ \left( \kappa^{H_5^0}_{WW} \right)^2 
				- \left( \kappa^{H_5^{++}}_{WW} \right)^2 \right) \left( s + t \right) 
			\right. \nonumber \\
			&& \left. \qquad \ \ 
				+ \, 2 \left( \kappa^{h^0}_{WW} \right)^2 m_{h^0}^2 
				+ 2 \left( \kappa^{H^0}_{WW} \right)^2 m_{H^0}^2
				+ 2 \left( \kappa^{H_5^0}_{WW} \right)^2 m_5^2
				+ \left( \kappa^{H_5^{++}}_{WW} \right)^2 m_5^2 \right],
	\label{eq:WWGM}
\end{eqnarray}
where I already used $u \simeq -(s + t)$.

As usual, we need the first line to equal $- \frac{1}{v^2} (s + t)$ in order to cancel the $\mathcal{O}(E^2/v^2)$ part of the gauge-boson-only diagrams---this imposes our first sum rule in the GM model,
\begin{equation}
	\boxed{
		\left( \kappa^{h^0}_{WW} \right)^2 + \left( \kappa^{H^0}_{WW} \right)^2
		+ \left( \kappa^{H_5^0}_{WW} \right)^2 - \left( \kappa^{H_5^{++}}_{WW} \right)^2
		= 1.
	}
\end{equation}
In terms of the actual values of the coupling modification factors in the GM model, we have
\begin{equation}
	c_H^2 + \frac{8}{3} s_H^2 + \left( \frac{1}{\sqrt{3}} s_H \right)^2 - \left( \sqrt{2} s_H \right)^2
		= c_H^2 + s_H^2 = 1. \qquad \qquad {\rm Success!}
\end{equation}

Doing the same thing for $W^+_L W^-_L \to Z_L Z_L$, for which now $h^0$, $H^0$, and $H_5^0$ contribute in the $s$ channel and $H_5^+$ contributes in the $t$ and $u$ channels, we get the matrix element,
\begin{eqnarray}
	\mathcal{M} \cdot \frac{1}{\sqrt{2}} &=& - \frac{1}{v^2} \cdot \frac{1}{\sqrt{2}} \left[ 
		\left( \kappa^{h^0}_{WW} \kappa^{h^0}_{ZZ} \right) \left( \frac{s^2}{s - m_{h^0}^2} \right)
		+ \left( \kappa^{H^0}_{WW} \kappa^{H^0}_{ZZ} \right) \left( \frac{s^2}{s - m_{H^0}^2} \right)
		+ \left( \kappa^{H_5^0}_{WW} \kappa^{H_5^0}_{ZZ} \right) \left( \frac{s^2}{s - m_5^2} \right)
		\right. \nonumber \\
		&& \left. \qquad \qquad \ \ \ 
		+ \left( \kappa^{H_5^+}_{WZ} \right)^2 
			\left( \frac{t^2}{t - m_5^2} + \frac{u^2}{u - m_5^2} \right) \right] \\
		&\simeq& - \frac{1}{v^2} \cdot \frac{1}{\sqrt{2}} \left[
			\left( \left( \kappa^{h^0}_{VV} \right)^2 + \left( \kappa^{H^0}_{VV} \right)^2
			+ \kappa^{H_5^0}_{WW} \kappa^{H_5^0}_{ZZ} 
			- \left( \kappa^{H_5^+}_{WZ} \right)^2 \right) s \right. \nonumber \\
		&& \left. \qquad \qquad \ \ \
			+ \left( \kappa^{h^0}_{VV} \right)^2 m_{h^0}^2 
			+ \left( \kappa^{H^0}_{VV} \right)^2 m_{H^0}^2
			+ \kappa^{H_5^0}_{WW} \kappa^{H_5^0}_{ZZ} m_5^2
			+ 2 \left( \kappa^{H_5^+}_{WZ} \right)^2 m_5^2 \right].
\end{eqnarray}
Here we have used the fact that the $\varphi WW$ and $\varphi ZZ$ coupling modification factors are the same for $h^0$ and $H^0$ but different for $H_5^0$.
We need the first line to equal $- \frac{1}{\sqrt{2} v^2} s$ in order to cancel the $\mathcal{O}(E^2/v^2)$ part of the gauge-boson-only diagrams, which gives us our second sum rule in the GM model:
\begin{equation}
	\boxed{
		\left( \kappa^{h^0}_{VV} \right)^2 + \left( \kappa^{H^0}_{VV} \right)^2
		+ \kappa^{H_5^0}_{WW} \kappa^{H_5^0}_{ZZ}
		- \left( \kappa^{H_5^+}_{WZ} \right)^2 = 1.
	}
\end{equation}
In terms of the actual values of the coupling modification factors in the GM model, we have
\begin{equation}
	c_H^2 + \frac{8}{3} s_H^2 + \left( \frac{1}{\sqrt{3}} s_H \right) \left( - \frac{2}{\sqrt{3}} s_H \right)
	- \left( -s_H \right)^2 
	= c_H^2 + s_H^2 = 1. 
\end{equation}
We see that even the opposite signs of the $H_5^0 WW$ and $H_5^0 ZZ$ couplings are important for the GM model to hang together properly.

Finally we can use the pieces of the matrix elements that do not grow with energy to get bounds on combinations of the scalar masses and couplings.  From $W^+_L W^-_L \to W^+_L W^-_L$ scattering, setting the second line of Eq.~(\ref{eq:WWGM}) equal to $16 \pi a_0$, taking $| {\rm Re} \, a_0| \leq 1/2$, dialing $\alpha$ so that $\kappa^{H^0}_{WW} = 0$ for the most conservative bound, and solving for $s_H$ yields
\begin{equation}
	s_H^2 \leq \frac{3}{2} \frac{(8 \pi v^2 - 2 m_{h^0}^2)}{(4 m_5^2 + 5 m_h^2)} 
	\simeq \left( \frac{755~{\rm GeV}}{m_5} \right)^2.
\end{equation}
Again we see a decoupling behaviour for the ``exotic'' triplet vev, which directly controls the couplings of the custodial-fiveplet states to gauge boson pairs and must fall with increasing $m_5$ at least as fast as
\begin{equation}
	\sin^2\theta_H \sim \frac{{\rm (number)} \cdot v^2}{m_5^2}.
\end{equation}
I \emph{have} worked through the details of decoupling in the GM model~\cite{Hartling:2014zca}, and the decoupling does indeed go with this power of $m_5$ (rather than faster, as in the 2HDM).

\section{Decoupling and alignment in the 2HDM}
\label{sec:decoup}

To round out these lectures I want to change gears and show you some details of the decoupling limit in the (CP-conserving) 2HDM as derived in full detail from the scalar potential.  I'll also discuss a different limit, that of so-called \emph{alignment without decoupling}, which can occur in the 2HDM.  The discussion in this section closely follows the classic papers Ref.~\cite{Gunion:2002zf} (on decoupling) and Ref.~\cite{Bernon:2015qea} (on alignment).

\subsection{Decoupling in the 2HDM}

The basic idea behind the decoupling limit is that, in this limit, the masses of all the new Higgs bosons become heavy while a single CP-even neutral Higgs boson ($h^0$) remains light and all of the couplings of $h^0$ to SM particles approach their SM values.  Not all extended Higgs models possess a decoupling limit (see homework question 4 for an example), but those that do are generally able to evade experimental constraints so long as the data remains consistent with the SM.\footnote{This is simultaneously useful and frustrating.}

Let's study decoupling in the CP-conserving 2HDM (see also Sec.~\ref{sec:2HDMdefs} for the notation used here).  The most general scalar potential can be written as\footnote{Here h.c. means Hermitian conjugate.}
\begin{eqnarray}
	V &=& m_{11}^2 \Phi_1^{\dagger} \Phi_1 + m_{22}^2 \Phi_2^{\dagger} \Phi_2
		- \left[ m_{12}^2 \Phi_1^{\dagger} \Phi_2 + {\rm h.c.} \right] \nonumber \\
	&& + \frac{1}{2} \lambda_1 \left( \Phi_1^{\dagger} \Phi_1 \right)^2
		+ \frac{1}{2} \lambda_2 \left( \Phi_2^{\dagger} \Phi_2 \right)^2
		+ \lambda_3 \left( \Phi_1^{\dagger} \Phi_1 \right) \left( \Phi_2^{\dagger} \Phi_2 \right) 
		+ \lambda_4 \left( \Phi_1^{\dagger} \Phi_2 \right) \left( \Phi_2^{\dagger} \Phi_1 \right)
		\nonumber \\
	&&
		+ \left\{ \frac{1}{2} \lambda_5 \left( \Phi_1^{\dagger} \Phi_2 \right)^2
			+ \left[ \lambda_6 \left( \Phi_1^{\dagger} \Phi_1 \right)
				+ \lambda_7 \left( \Phi_2^{\dagger} \Phi_2 \right) \right]
				\left( \Phi_1^{\dagger} \Phi_2 \right) + {\rm h.c.} \right\}.
	\label{eq:V}
\end{eqnarray}		
Here $m_{12}^2$, $\lambda_5$, $\lambda_6$, and $\lambda_7$ can be complex in general; we will choose them real, thereby avoiding explicit CP violation.\footnote{Often you'll see the 2HDM scalar potential written down with $\lambda_6 = \lambda_7 = 0$.  This is because \emph{natural flavour conservation}~\cite{Glashow:1976nt,Paschos:1976ay} can be achieved in the 2HDM by imposing a discrete $Z_2$ symmetry $\Phi_1 \to -\Phi_1$ (along with, e.g., $d_R \to -d_R$ and $\ell_R \to -\ell_R$, which yields the so-called Type-II 2HDM).  This automatically prevents tree-level flavour-changing neutral Higgs interactions.  The discrete symmetry is usually allowed to be softly broken by $m_{12}^2 \neq 0$.}
Recall that the vevs are
\begin{equation}
	\langle \Phi_1 \rangle = \left( \begin{array}{c} 0 \\ v_1/\sqrt{2} \end{array} \right), \qquad \qquad
	\langle \Phi_2 \rangle = \left( \begin{array}{c} 0 \\ v_2/\sqrt{2} \end{array} \right),
\end{equation}
and the $W$ mass is $M_W^2 = g^2 (v_1^2 + v_2^2)/4 = g^2 v^2/4$, where $v \simeq 246~{\rm GeV}$ is the Higgs vev in the SM.  We also define $\tan\beta \equiv v_2/v_1$.

The dimensionful parameters $m_{11}^2$ and $m_{22}^2$ can be eliminated using the minimization conditions $\partial V/\partial v_1 = 0$ and $\partial V/\partial v_2 = 0$, to give
\begin{eqnarray}
	m_{11}^2 &=& m_{12}^2 \tan\beta - \frac{1}{2} v^2 \left[ \lambda_1 \cos^2\beta + \lambda_{345} \sin^2\beta + 3 \lambda_6 \sin\beta \cos\beta + \lambda_7 \sin^2\beta \tan\beta \right], \\
	m_{22}^2 &=& m_{12}^2 \cot\beta - \frac{1}{2} v^2 \left[ \lambda_2 \sin^2\beta + \lambda_{345} \cos^2\beta + \lambda_6 \cos^2\beta \cot\beta + 3 \lambda_7 \sin\beta \cos\beta \right],
\end{eqnarray}
where we also defined the combination $\lambda_{345} \equiv \lambda_3 + \lambda_4 + \lambda_5$.

We started with three dimensionful parameters in $V$: $m_{11}^2$, $m_{22}^2$, and $m_{12}^2$.  We have now eliminated two of them in favour of the vevs $v_1$ and $v_2$, which are each \emph{bounded from above} by $\sqrt{v_1^2 + v_2^2} = v \simeq 246~{\rm GeV}$.  There is one more remaining dimensionful parameter, $m_{12}^2$, which we will see in a minute controls the heavy Higgs boson masses and can be taken arbitrarily large, thereby implementing the decoupling limit.\footnote{If on the other hand we were to \emph{eliminate} $m_{12}^2$ by imposing an exact $Z_2$ symmetry $\Phi_1 \to -\Phi_1$ on the potential, then \emph{all} the Higgs masses turn out to be bounded by requiring perturbativity of the couplings $\lambda_i$ (or technically, by requiring perturbative unitarity of $\varphi_1 \varphi_2 \to \varphi_3 \varphi_4$ scattering amplitudes) to be below about 700~GeV, and there is no decoupling limit.  The bounds on the Higgs masses in the 2HDM in this scenario were worked out in Ref.~\cite{Kanemura:1993hm}. \label{fn:exactZ2}}

The physical spectrum is as follows.
There is one CP-odd neutral Higgs boson $A^0$,
	\begin{equation}
		A^0 = -\sin\beta \, \phi_1^{0,i} + \cos\beta \, \phi_2^{0,i},
	\end{equation}
	with mass
	\begin{equation}
		m_{A^0}^2 = \frac{m_{12}^2}{\sin\beta \cos\beta} 
		- \frac{1}{2} v^2 \left( 2 \lambda_5 + \lambda_6 \cot\beta + \lambda_7 \tan\beta \right).
	\end{equation}
	Notice that large $m_{12}^2$ directly translates into large $m_{A^0}^2$.
	
There is also one charged Higgs pair $H^{\pm}$,
	\begin{equation}
		H^{\pm} = -\sin\beta \, \phi_1^{\pm} + \cos\beta \, \phi_2^{\pm},
	\end{equation}
	with mass
	\begin{equation}
		m_{H^{\pm}}^2 = m_{A^0}^2 + \frac{1}{2} v^2 \left( \lambda_5 - \lambda_4 \right).
	\end{equation}
	Notice that $m_{H^{\pm}}^2$ tracks $m_{A^0}^2$ at large mass.\footnote{In particular, $m_{H^{\pm}}^2 - m_{A^0}^2 \sim \mathcal{O}(\lambda_i v^2)$ implies that $m_{H^{\pm}} - m_{A^0} \sim \mathcal{O}(\lambda_i v^2/m_{A^0})$, so the actual mass splitting gets smaller and smaller as $m_{A^0}$ increases.  The same holds for the mass splitting between $H^0$ and $H^{\pm}$.  This ensures that the additional Higgs bosons' contribution to electroweak precision observables (particularly the $T$ parameter) become increasingly suppressed in the decoupling limit.}
	
Finally there are two CP-even neutral Higgs bosons $h^0$ and $H^0$, whose mass-squared matrix is given in the basis $( \phi_1^{0,r}, \phi_2^{0,r} )$ by
	\begin{equation}
		M^2 = m_{A^0}^2 \left( \begin{array}{cc} \sin^2\beta & -\sin\beta \cos\beta \\
						-\sin\beta \cos\beta & \cos^2\beta \end{array} \right) + B^2,
	\end{equation}
	where
	\begin{equation}
		B^2 = v^2 \left( \begin{array}{cc}
			\lambda_1 \cos^2\beta + 2 \lambda_6 \sin\beta \cos\beta + \lambda_5 \sin^2 \beta 
			& (\lambda_3 + \lambda_4) \sin\beta \cos\beta + \lambda_6 \cos^2\beta 
				+ \lambda_7 \sin^2 \beta \\
			(\lambda_3 + \lambda_4) \sin\beta \cos\beta + \lambda_6 \cos^2\beta 
				+ \lambda_7 \sin^2 \beta
			& \lambda_2 \sin^2\beta + 2 \lambda_7 \sin\beta \cos\beta + \lambda_5 \cos^2\beta
			\end{array} \right).
	\end{equation}
	In the absence of the $B^2$ part (i.e., if we were to ignore pieces of order $\lambda_i v^2$ compared to those of order $m_{A^0}^2$), the mass eigenstates become
	\begin{eqnarray}
		h^0 &=& \cos\beta \, \phi_1^{0,r} + \sin\beta \, \phi_2^{0,r}
			\qquad \qquad \ \ \  {\rm with \ mass \ zero}, \\
		H^0 &=& -\sin\beta \, \phi_1^{0,r} + \cos\beta \, \phi_2^{0,r} 
			\qquad \qquad {\rm with \ mass} \ m_{A^0}. 
	\end{eqnarray}
	Notice that the mixing angles that define $H^0$ in this approximation are the \emph{same} as those for $A^0$ and $H^{\pm}$.  Keeping the $\mathcal{O}(\lambda_i v^2)$ pieces, the actual mixing angle definition for the CP-even scalars is\footnote{Note the rather unfortunate historical choice of convention.  This is too well established to deviate from.}
	\begin{eqnarray}
		h^0 &=& -\sin\alpha \, \phi_1^{0,r} + \cos\alpha \, \phi_2^{0,r}, \\
		H^0 &=& \cos\alpha \, \phi_1^{0,r} + \sin\alpha \, \phi_2^{0,r}.
	\end{eqnarray}
	This means that the \emph{mismatch} between the $h^0$--$H^0$ mixing angle and that of $G^0$--$A^0$ and $G^{\pm}$--$H^{\pm}$ is parameterized by $\cos(\beta-\alpha)$, which goes to zero in the decoupling limit.  
	
An exact (though not necessarily the most useful) way to write this in terms of the elements of the matrix $B^2$ is
\begin{eqnarray}
	&& \qquad \qquad \cos^2(\beta - \alpha) = \frac{m_L^2 - m_{h^0}^2}{m_{H^0}^2 - m_{h^0}^2}, 
	\qquad {\rm where} \\
	&& m_L^2 = B_{11}^2 \cos^2\beta + B_{22}^2 \sin^2\beta + 2 B_{12}^2 \sin\beta \cos\beta.
\end{eqnarray}
In the decoupling limit, $m_{H^0} \sim m_{A^0} \sim m_{H^{\pm}} \gg m_{h^0}, v$, and $\cos(\beta-\alpha)$ becomes small.  But in that limit, a cancellation is also happening in the numerator of the above expression.  In the decoupling limit one obtains the approximate relation\footnote{Notice that this expression is for $\cos(\beta-\alpha)$, not $\cos^2(\beta-\alpha)$!}
\begin{equation}
	\cos(\beta - \alpha) \simeq \frac{\hat \lambda v^2}{m_{A^0}^2 - \lambda_A v^2}
		\simeq \frac{\hat \lambda v^2}{m_{H^0}^2 -  m_{h^0}^2},
	\label{eq:cba}
\end{equation}
where $\hat\lambda$ and $\lambda_A$ are some combinations of the $\lambda_i$.\footnote{The expressions for $\hat \lambda$ and $\lambda_A$ are~\cite{Gunion:2002zf}
\begin{eqnarray}
	\hat \lambda &=& \sin\beta \cos\beta \left[ \lambda_1 \cos^2\beta - \lambda_2 \sin^2\beta 
		- \lambda_{345} \cos(2\beta) \right] - \lambda_6 \cos\beta \cos(3\beta)
		- \lambda_7 \sin\beta \sin(3\beta), \\
	\lambda_A &=& \cos(2\beta) \left[ \lambda_1 \cos^2\beta - \lambda_2 \sin^2\beta \right]
		+ \lambda_{345} \sin^2(2 \beta) - \lambda_5 + 2 \lambda_6 \cos\beta \sin(3\beta)
		- 2 \lambda_7 \sin\beta \cos(3\beta).
\end{eqnarray}
The important thing here is that they are combinations of dimensionless quartic-scalar couplings which must not be too big or perturbation theory will break down.
}
In other words, rather than $\cos^2(\beta-\alpha)$ going like $\mathcal{O}(v^2/m_{A^0}^2)$, as we found in Sec.~\ref{sec:2HDM} was required to satisfy perturbative unitarity of longitudinal vector boson scattering, we find instead that the 2HDM \emph{decouples faster than strictly necessary}, with $\cos^2(\beta-\alpha)$ going like $\mathcal{O}(v^4/m_{A^0}^4)$.
This matters a lot when we want to go looking for deviations in the couplings of the 125~GeV Higgs boson to SM particles, which could constitute signs of non-minimality in the Higgs sector.

\subsection{Implications for Higgs coupling measurements}

Measuring the couplings of the already-discovered 125~GeV Higgs boson is a good way to search for new physics because these couplings generally get modified from their SM values in extended Higgs sectors.  The maximum allowed size of these modifications is linked, via perturbative unitarity of longitudinal gauge boson scattering (or alternatively, the decoupling behaviour of the theory), to the energy scale at which the additional Higgs bosons must appear, which is extremely useful because detecting a deviation points us directly to the scale of the new physics.

Let's see how this plays out in the couplings of $h^0$ to SM particles in the 2HDM when we are near the decoupling limit.  This analysis also gives us insight into which couplings are most likely to show deviations (or in other words, what the expected pattern of deviations is) in this particular model.

First consider the $h^0$ couplings to $WW$ and $ZZ$.  We have,
\begin{equation}
	h^0 V_{\mu} V_{\nu}: \ \frac{2 i M_V^2}{v} g_{\mu\nu} \kappa^{h^0}_{VV}
\end{equation}
where
\begin{equation}
	\kappa^{h^0}_{VV} = \sin(\beta-\alpha) = \sqrt{1 - \cos^2(\beta-\alpha)}
		\simeq 1 - \frac{1}{2} \cos^2(\beta-\alpha),
\end{equation}
where in the last step we made an expansion of the square root for $\cos(\beta-\alpha) \ll 1$.  Since $\cos^2(\beta-\alpha) \sim \mathcal{O}(v^4/m_{A^0}^4)$ in the decoupling limit, we see that the deviation of the $h^0VV$ coupling from its SM value is of this order, i.e.,
\begin{equation}
	\kappa^{h^0}_{VV} = 1 - \mathcal{O}(v^4/m_{A^0}^4).
\end{equation}

In order to get a sense of scale, let's compare this to the behaviour of the $h^0$ couplings to fermion pairs (for concreteness I'll work in the Type II 2HDM; for full details see the 2HDM review article~\cite{Branco:2011iw}).  These are given by
\begin{equation}
	h^0 f \bar f: \ -i \frac{m_f}{v} \kappa^{h^0}_f
\end{equation}
where for up-type fermions (e.g., $t \bar t$ and $c \bar c$), 
\begin{equation}
	\kappa^{h^0}_t = \frac{\cos\alpha}{\sin\beta} = \sin(\beta-\alpha) + \cot\beta \cos(\beta-\alpha)
		\simeq 1 + \cot\beta \cos(\beta-\alpha),
\end{equation}
where again the approximation is an expansion to leading nontrivial order in small $\cos(\beta-\alpha)$.  Thus we find that the deviation of the $h^0 t \bar t$ and $h^0 c \bar c$ couplings from their SM values are of order $\cos(\beta-\alpha)$, in contrast to that of the $h^0 VV$ coupling which goes like $\cos^2(\beta-\alpha)$.  In particular,
\begin{equation}
	\kappa^{h^0}_t = 1 + \mathcal{O}(\cot\beta \cdot v^2/m_{A^0}^2).
\end{equation}
This has important implications for Higgs coupling measurements in the 2HDM: near the decoupling limit, we expect the deviation in the $h^0$ coupling to up-type quarks to be significantly larger than the deviation in the coupling to vector boson pairs, unless $\cot\beta$ is very small.\footnote{Perturbativity of the top quark Yukawa coupling in this model prevents $\cot\beta$ from being larger than about 2 or 3.}

Let's do the same for $h^0$ couplings to down-type fermions ($b \bar b$, $\tau^+\tau^-$, etc.).  We have,
\begin{equation}
	\kappa^{h^0}_b = - \frac{\sin\alpha}{\cos\beta} = \sin(\beta-\alpha) - \tan\beta \cos(\beta - \alpha)
		\simeq 1 - \tan\beta \cos(\beta-\alpha).
\end{equation}
This deviation,
\begin{equation}
	\kappa^{h^0}_b = 1 - \mathcal{O}(\tan\beta \cdot v^2/m_{A^0}^2),
\end{equation}
is even more exciting, because not only does it decouple much more slowly than the deviation in the $h^0VV$ coupling, it is also enhanced by $\tan \beta$, which can be as large as $\mathcal{O}(50)$!  This enhancement of the deviation in the $h^0 b \bar b$ and $h^0 \tau \tau$ couplings at large $\tan\beta$ is known as ``delayed decoupling'' and has been studied in some detail in the literature (including taking into account one-loop corrections to the Higgs couplings), particularly in the Minimal Supersymmetric Standard Model (MSSM) whose Higgs sector is a Type-II 2HDM.

The take-home message is that, if you think that you're dealing with a Type-II 2HDM with relatively heavy additional Higgs bosons, the couplings of $h^0$ in which you should expect the largest deviations from the SM are $h^0 b \bar b$ and $h^0 \tau\tau$.  

Meanwhile, the couplings of the other (heavy) CP-even neutral Higgs boson $H^0$ go like
\begin{eqnarray}
	\kappa^{H^0}_{VV} &=& \cos(\beta - \alpha) \simeq \mathcal{O}(v^2/m_{A^0}^2), \\
	\kappa^{H^0}_t &=& \frac{\sin\alpha}{\sin\beta} = \cos(\beta - \alpha) - \cot\beta \sin(\beta - \alpha)
		\simeq - \cot\beta, \\
	\kappa^{H^0}_b &=& \frac{\cos\alpha}{\cos\beta} 
		= \cos(\beta - \alpha) + \tan\beta \sin(\beta-\alpha) \simeq \tan\beta;
\end{eqnarray}
in particular, while its production through its coupling to vector boson pairs (e.g., through vector boson fusion) is strongly suppressed, its production through its coupling to down-type fermions (e.g., in association with a $b \bar b$ pair) can be strongly enhanced at large $\tan \beta$.  This latter process is the basis for very strong LHC limits on $H^0$ (and $A^0$) in the MSSM.

Finally I want to say a few words about triple-Higgs couplings.  The $h^0$ self-coupling (a major target of the High-Luminosity LHC physics program) is~\cite{Gunion:2002zf}
\begin{equation}
	h^0 h^0 h^0: \ i g_{h^0h^0h^0} \simeq -\frac{3 i m_{h^0}^2}{v} + 6 i \hat\lambda v \cos(\beta-\alpha),
\end{equation}
where again the approximation is for small $\cos(\beta-\alpha)$, the first term is the SM result, $\hat \lambda$ is the same combination of quartic scalar couplings that appears in Eq.~(\ref{eq:cba}), and the second term (the deviation) is $\mathcal{O}(v^2/m_{A^0}^2)$ relative to the first term.

The other experimentally-relevant triple-Higgs coupling is that of $h^0$ to $H^+H^-$, because it controls the charged Higgs contribution to the $h^0 \to \gamma\gamma$ loop.  Near the decoupling limit it can be written as~\cite{Gunion:2002zf}
\begin{equation}
	h^0 H^+ H^-: \ i g_{h^0 H^+ H^-} \simeq - v(\lambda_F + \lambda_T) 
	+ \mathcal{O}\left( \cos(\beta-\alpha) \right),
\end{equation}
where $\lambda_F$ and $\lambda_T$ are some messy combinations of the $\lambda_i$ that can be found in Eqs.~(28) and (52) of Ref.~\cite{Gunion:2002zf}.  The important point is that $g_{h^0 H^+H^-}$ is some combination of quartic couplings times $v$: in particular, it does \emph{not} grow with $m_{H^{\pm}}$.  This means that, once it is combined with the calculation of the one-loop integrals for the $H^{\pm}$ contribution to $h^0 \to \gamma\gamma$, which themselves go like $m_{h^0}^2/m_{H^{\pm}}^2$ for $m_{H^{\pm}} \gg m_{h^0}$, the charged Higgs loop contribution to the effective $h^0 \gamma\gamma$ coupling gives a modification that goes like $\mathcal{O}(v^2/m_{A^0}^2)$.

The bottom line is that when $m_{A^0} \gg v$, we also have $m_{H^0} \sim m_{H^{\pm}} \sim m_{A^0}$ and the couplings of $h^0$ to SM particles deviate from their SM values by \emph{at most} a factor of $(1 + \mathcal{O}(v^2/m_{A^0}^2))$.

The discovered Higgs boson's best-measured couplings have (at the time of writing) been measured to about the $\pm 10\%$ level, and are currently consistent with the SM predictions.  Does that mean that we must have
\begin{equation}
	m_{A^0} > \frac{1}{\sqrt{0.1}} v \sim 3 v?
\end{equation}
Not necessarily!  Read on for details.

\subsection{SM-like Higgs couplings without decoupling: the alignment limit}

To better understand the Higgs couplings and the decoupling limit in the 2HDM, it's useful to do a change of basis among the two doublets to the so-called \emph{Higgs basis}.

First consider the situation in which $(\beta - \alpha) = \pi/2$, so that $\sin(\beta - \alpha) = 1$ and $\cos(\beta - \alpha) = 0$.  The composition of the lighter Higgs boson is then
\begin{eqnarray}
	h^0 &=& -\sin\alpha \, \phi_1^{0,r} + \cos\alpha \, \phi_2^{0,r} \\
	&=& -\sin(\beta - \pi/2) \, \phi_1^{0,r} + \cos(\beta - \pi/2) \, \phi_2^{0,r} \\
	&=& \cos\beta \, \phi_1^{0,r} + \sin\beta \, \phi_2^{0,r}.
\end{eqnarray}
Compare this to the composition of the Goldstone bosons,
\begin{eqnarray}
	G^0 &=& \cos\beta \, \phi_1^{0,i} + \sin\beta \, \phi_2^{0,i}, \\
	G^+ &=& \cos\beta \, \phi_1^+ + \sin\beta \, \phi_2^+,
\end{eqnarray}
and the ``direction'' in field space of the vev:
\begin{equation}
	v = \cos\beta \, v_1 + \sin\beta \, v_2,
\end{equation}
where we used $\cos\beta = v_1/v$, $\sin\beta = v_2/v$, and $v_1^2 + v_2^2 = v^2$.

Clearly in this special case of $\alpha = \beta - \pi/2$ we can simply rotate the entire doublets $\Phi_1$ and $\Phi_2$ by an angle $\beta$ to define a new basis:
\begin{eqnarray}
	\Phi_H &\equiv& \cos\beta \, \Phi_1 + \sin\beta \, \Phi_2 
		= \left( \begin{array}{c} G^+ \\ (v + h^0 + i G^0)/\sqrt{2} \end{array} \right), \\
	\Phi_0 &\equiv& -\sin\beta \, \Phi_1 + \cos\beta \, \Phi_2
		= \left( \begin{array}{c} H^+ \\ (H^0 + i A^0)/\sqrt{2} \end{array} \right).
\end{eqnarray}
This basis is known as the \emph{Higgs basis} and is defined by the vev being all in one doublet $\Phi_H$.  Notice that the Goldstone bosons always live in $\Phi_H$ while $H^+$ and $A^0$ always live in $\Phi_0$.

When $(\beta - \alpha) \neq \pi/2$, we have to take into account the mixing of $h^0$ and $H^0$ between the two doublets in the Higgs basis.  In this unrestricted case we have,
\begin{eqnarray}
	\Phi_H &=& \left( \begin{array}{cc} G^+ \\ 
		(v + h^0 \sin(\beta - \alpha) + H^0 \cos(\beta - \alpha) + i G^0)/\sqrt{2} \end{array} \right), \\
	\Phi_0 &=& \left( \begin{array}{cc} H^+ \\
		( - h^0 \cos(\beta - \alpha) + H^0 \sin(\beta - \alpha) + i A^0)/\sqrt{2} \end{array} \right).
\end{eqnarray}
Decoupling forces $\cos(\beta - \alpha) \to 0$; in this limit, $h^0$ lives entirely in $\Phi_H$ and $H^0$ lives entirely in $\Phi_0$.  The upshot of this is that, when $\cos(\beta - \alpha) = 0$, $\Phi_H$ plays the role of the SM Higgs doublet, while $\Phi_0$ (with no vev) is essentially just an extra SU(2) doublet of scalar particles, which can and typically do couple to fermions but which have nothing to do with the Higgs mechanism and can be taken arbitrarily heavy without causing any trouble with the unitarization of longitudinal vector boson scattering amplitudes.  

But we can of course also have $\cos(\beta - \alpha) \to 0$ \emph{without} making the scalars in $\Phi_0$ very heavy: this is called the \emph{alignment limit} or \emph{alignment without decoupling} (because the mixing angle of $h^0$ and $H^0$ is ``aligned'' with that of the vev and the Goldstones).  The alignment limit is interesting because in this limit $h^0$ can have very SM-like couplings and yet we can also have additional light new scalars at the weak scale.

To see how the alignment limit can arise from the scalar potential of the 2HDM, we first rewrite the scalar potential in the Higgs basis as~\cite{Bernon:2015qea}
\begin{eqnarray}
	V &=& Y_1 \Phi_H^{\dagger} \Phi_H + Y_2 \Phi_0^{\dagger} \Phi_0
		+ Y_3 \left( \Phi_H^{\dagger} \Phi_0 + {\rm h.c.} \right) \nonumber \\
	&& + \frac{1}{2} Z_1 \left( \Phi_H^{\dagger} \Phi_H \right)^2
		+ \frac{1}{2} Z_2 \left( \Phi_0^{\dagger} \Phi_0 \right)^2
		+ Z_3 \left( \Phi_H^{\dagger} \Phi_H \right) \left( \Phi_0^{\dagger} \Phi_0 \right)
		+ Z_4 \left( \Phi_H^{\dagger} \Phi_0 \right) \left( \Phi_0^{\dagger} \Phi_H \right) \nonumber \\
	&& + \left\{ \frac{1}{2} Z_5 \left( \Phi_H^{\dagger} \Phi_0 \right)^2
		+ \left[ Z_6 \left( \Phi_H^{\dagger} \Phi_H \right)
			+ Z_7 \left( \Phi_0^{\dagger} \Phi_0 \right) \right] \left( \Phi_H^{\dagger} \Phi_0 \right)
		+ {\rm h.c.} \right\}.
\end{eqnarray}
This is just a change of basis of our original potential in Eq.~(\ref{eq:V}), so the $Y_i$ and $Z_i$ are just linear combinations of our original $m_{ij}^2$ and $\lambda_i$, respectively.  Taking $\lambda_6 = \lambda_7 = 0$ in the original potential (for simplicity, and so that we can implement natural flavour conservation using a $\Phi_1 \to - \Phi_1$ symmetry), we have~\cite{Bernon:2015qea}
\begin{eqnarray}
	Y_1 &=& m_{11}^2 \cos^2\beta + m_{22}^2 \sin^2\beta - m_{12}^2 \sin(2\beta), \\
	Y_2 &=& m_{11}^2 \sin^2\beta + m_{22}^2 \cos^2\beta + m_{12}^2 \sin(2\beta), \\
	Y_3 &=& \frac{1}{2} (m_{22}^2 - m_{11}^2) \sin(2\beta) - m_{12}^2 \cos(2\beta), \\
	Z_1 &=& \lambda_1 \cos^4\beta + \lambda_2 \sin^4\beta 
		+ \frac{1}{2} \lambda_{345} \sin^2(2\beta), 
		\qquad \qquad {\rm with} \quad \lambda_{345} = \lambda_3 + \lambda_4 + \lambda_5 \\
	Z_2 &=& \lambda_1 \sin^4\beta + \lambda_2 \cos^4\beta 
		+ \frac{1}{2} \lambda_{345} \sin^2(2\beta), \\
	Z_i &=& \frac{1}{4} \sin^2(2\beta) \left[ \lambda_1 + \lambda_2 - 2 \lambda_{345} \right] 
		+ \lambda_i,  \qquad \qquad \qquad {\rm for} \quad i = 3, 4, 5 \\
	Z_6 &=& -\frac{1}{2} \sin(2\beta) \left[ \lambda_1 \cos^2\beta - \lambda_2 \sin^2\beta 
		- \lambda_{345} \cos(2\beta) \right], \\
	Z_7 &=& -\frac{1}{2} \sin(2\beta) \left[ \lambda_1 \sin^2\beta - \lambda_2 \cos^2\beta
		+ \lambda_{345} \cos(2\beta) \right].
\end{eqnarray}
The mass-squared parameters $Y_1$ and $Y_3$ can be eliminated using the scalar potential minimization conditions:
\begin{equation}
	Y_1 = -\frac{1}{2} Z_1 v^2, \qquad \qquad \qquad Y_3 = -\frac{1}{2} Z_6 v^2.
\end{equation}
This leaves the mass-squared parameter $Y_2$ as a free parameter, and the masses of the states in $\Phi_0$ are mainly controlled by $Y_2$ when $Y_2 \gg v^2$.

In terms of this parameterization, the physical scalar masses are given by,
\begin{eqnarray}
	m_{H^{\pm}}^2 &=& Y_2 + \frac{1}{2} Z_3 v^2, \\
	m_{A^0}^2 &=& Y_2 + \frac{1}{2} (Z_3 + Z_4 - Z_5) v^2, \\
	M_{h^0,H^0}^2 &=& \left( \begin{array}{cc} Z_1 v^2 & Z_6 v^2 \\
						Z_6 v^2 & m_{A^0}^2 + Z_5 v^2 \end{array} \right),
\end{eqnarray}
where the mass-squared matrix for $h^0$ and $H^0$ is written in the basis $(\phi_H^{0,r}, \phi_0^{0,r})$.  So simple!

It becomes very clear that there are \emph{two} ways to make the mixing angle in $M_{h^0,H^0}^2$ be small (so that $h^0$ lives mostly in $\Phi_H$ and has SM-like couplings):
\begin{enumerate}
\item make $m_{A^0}^2 \gg v^2$ so that the $(2,2)$ element becomes large; or
\item make $Z_6 \ll 1$ so that the off-diagonal elements become small.
\end{enumerate}
The first choice gives the decoupling limit.  The second choice gives \emph{alignment without decoupling}.  In either case the SM-like Higgs boson mass-squared is $m_{h^0} \simeq Z_1 v^2$.

Explicitly, we can write
\begin{equation}
	\cos(\beta - \alpha) = \frac{ - Z_6 v^2}{\sqrt{(m_{H^0}^2 - m_{h^0}^2)(m_{H^0}^2 - Z_1 v^2)}}.
\end{equation}
$Z_6 \to 0$ defines the alignment limit.

What are the implications for Higgs physics in the alignment limit?  For the $h^0VV$ and $h^0f \bar f$ couplings, only $\cos(\beta-\alpha)$ and $\tan\beta$ matter: there is no difference in the pattern of couplings between the alignment limit and the usual decoupling limit (at least at tree level).  

Differences do show up, however, in triple-Higgs couplings.  Let's define
\begin{equation}
	\cos(\beta - \alpha) = - \eta Z_6,
\end{equation}
where
\begin{equation}
	\eta = \frac{v^2}{\sqrt{(m_{H^0}^2 - m_{h^0}^2)(m_{H^0}^2 - Z_1 v^2)}} 
	= \left\{ \begin{array}{l} \mathcal{O}(1) \ {\rm when} \ m_{H^0}^2 \sim \mathcal{O}(v^2), \ {\rm or} \\
		\mathcal{O}(v^2/m_{H^0}^2) \ll 1 \ {\rm in \ the \ decoupling \ limit}. \end{array} \right.
\end{equation}

The $h^0 h^0 h^0$ coupling is $i g_{h^0h^0h^0}$ where~\cite{Bernon:2015qea}
\begin{eqnarray}
	g_{h^0h^0h^0} &=& -3 v \left[ Z_1 \sin^3(\beta-\alpha) 
		+ Z_{345} \sin(\beta-\alpha) \cos^2(\beta-\alpha) \right. \nonumber \\
		&& \left. \qquad + \, 3 Z_6 \cos(\beta-\alpha) \sin^2(\beta-\alpha) 
		+ Z_7 \cos^3(\beta-\alpha) \right],
\end{eqnarray}
where $Z_{345} = Z_3 + Z_4 + Z_5$.  We can do an expansion in the decoupling limit when $\eta \ll 1$:
\begin{equation}
	g_{h^0h^0h^0} = g_{hhh}^{\rm SM} 
		\left[ 1 - \frac{2 \eta Z_6^2}{Z_1} + \mathcal{O}(\eta^2 Z_6^2) \right], \qquad \qquad {\rm decoupling \ limit}
\end{equation}
where $g_{hhh}^{\rm SM} = -3 m_{h^0}^2/v$ is the coupling in the SM.  The deviation from the SM is always a suppression, and its size is $\mathcal{O}(\cos(\beta-\alpha))$.
Contrast this with the alignment limit, $\eta \sim \mathcal{O}(1)$ and $|Z_6| \ll 1$, in which~\cite{Bernon:2015qea}
\begin{eqnarray}
	g_{h^0h^0h^0} &=& g_{hhh}^{\rm SM} \left\{ 1 
		+ \left[ \left( Z_7 \tan(2\beta) - \frac{1}{2} Z_1 \right) \eta^2 - 2 \eta \right] \frac{Z_6^2}{Z_1}
		\right. \nonumber \\
		&& \left. \qquad
		+ \left( 2 \cot(2\beta) - \tan(2\beta) \right) \eta^2 \frac{Z_6^3}{Z_1} 
		+ \mathcal{O}(Z_6^3) \right\}, \qquad \qquad {\rm alignment \ limit} 
\end{eqnarray}
where we keep explicit only those $\mathcal{O}(Z_6^3)$ terms that could potentially be enhanced at large or small $\tan\beta$.  The deviation from the SM can be an enhancement or a suppression, and it is of order $Z_6^2 \sim \cos^2(\beta-\alpha)$.

The $H^0 h^0 h^0$ coupling (which is especially interesting for $m_{H^0} > 2 m_{h^0}$ because it can give rise to resonant $h^0h^0$ production) is likewise given by $i g_{H^0 h^0 h^0}$, where~\cite{Bernon:2015qea}
\begin{eqnarray}
	g_{H^0 h^0 h^0} &=& \frac{1}{v} \left\{ 3 Z_6 v^2 
		- \left[ m_{h^0}^2 - 4 Z_6 v^2 \cot(2\beta) 
			+ 2 (Z_6 - Z_7) v^2 \tan(2\beta) \right] \cos(\beta-\alpha) \right. \nonumber \\
		&& \qquad \left. + \, \mathcal{O}(\cos^2(\beta-\alpha)) \right\}.
\end{eqnarray}
In the decoupling limit with $Z_6 \sim \mathcal{O}(1)$, this coupling is unsuppressed:
\begin{equation}
	g_{H^0h^0h^0} \simeq 3 Z_6 v + \mathcal{O}(\cos(\beta-\alpha)),
\end{equation}
while in the alignment limit with $\eta \sim \mathcal{O}(1)$ it is suppressed:
\begin{equation}
	g_{H^0h^0h^0} \simeq \mathcal{O}(v \cos(\beta-\alpha)).
\end{equation}

Finally let's consider the $h^0 H^+ H^-$ coupling, which is important for the charged Higgs loop contribution to $h^0 \to \gamma\gamma$.  That coupling is given by
\begin{equation}
	g_{h^0H^+H^-} = -v \left[ Z_3 + \mathcal{O}(\cos(\beta-\alpha)) \right].
\end{equation}
It is \emph{not} suppressed in either the alignment limit or the decoupling limit, instead approaching a finite nonzero value controlled by the combination of quartic couplings $Z_3$.  I mentioned before that in the decoupling limit the loop integrals involving the charged Higgs give a suppression factor of $\mathcal{O}(m_{h^0}^2/m_{H^{\pm}}^2)$, so that the overall charged Higgs contribution to the $h^0 \to \gamma\gamma$ effective coupling also decouples.  But in the alignment limit we have $m_{H^{\pm}} \sim v$, so that this contribution can be important.  Aside from direct searches for the (light) additional scalars in the alignment limit, the measurement of $h^0 \to \gamma\gamma$ provides one of the most sensitive tests of this scenario.

As a final comment I should mention that $Z_6 \ll 1$ (necessary to implement the alignment limit) is a fine-tuning of the 2HDM parameters: within the original 2HDM, there is no symmetry reason why $Z_6 \to 0$ should be expected.  In contrast, large $Y_2$ (necessary to implement the decoupling limit) is in some ways more natural than $Y_2 \sim \mathcal{O}(v^2)$: if the 2HDM is an effective theory below some cutoff $\Lambda \gg v$, the hierarchy problem applies to each of the dimensionful parameters $Y_1$, $Y_2$ and $Y_3$, and letting one of them be rather large generically reduces the amount of fine-tuning.  That being said, certain extensions of the 2HDM allow for the introduction of new symmetries that can lead naturally to alignment without fine-tuning.\footnote{For a recent talk on this with lots of references, see Ref.~\cite{Haber:2022swy}.}

\section{Concluding remarks}
\label{sec:summary}

My objective in these lectures has been threefold:
\begin{itemize}
\item to demonstrate the power and versatility of perturbative unitarity techniques;
\item to illustrate the intricate and beautiful nature of the unitarity cancellations in extended Higgs sectors; and
\item to expose you to the structure of some exotic (and not-so-exotic) extended Higgs models.
\end{itemize}

I want to finish by emphasizing that not all extended Higgs sectors possess a decoupling limit.  Those that do not will still obey perturbative unitarity (so long as their couplings remain perturbative); they simply do not possess a limit in which the extra scalars can be taken very heavy and in which a light Higgs boson can be made to resemble the SM Higgs more and more precisely.  This implies that models that do not possess a decoupling limit can be definitively excluded (or discovered!) by sufficiently precise measurements of the SM-like Higgs boson's couplings.

\section{Homework questions}
\label{sec:homework}

\begin{enumerate}
\item The upper bound on the SM Higgs boson mass from perturbative unitarity of longitudinal vector boson scattering is ``actually'' an upper bound on the dimensionless quartic coupling $\lambda$ in the Higgs potential, $V = -\mu^2 \Phi^{\dagger} \Phi + \lambda (\Phi^{\dagger} \Phi)^2$ -- this is why its bound is related to the validity of perturbation theory.  Compute the Higgs mass in terms of $\lambda$ and $v$ and translate the bound on the Higgs mass calculated in these lectures, $m_h^2 \leq 8 \pi v^2/3 \simeq (712~{\rm GeV})^2$, into an upper bound on $\lambda$.  [Note that $v \simeq 246$~GeV is normalized such that $\langle \Phi \rangle = (0, v/\sqrt{2})^T$.]

\item Perform a calculation of the upper bound on the SM Higgs boson mass from perturbative unitarity by computing the zeroth partial wave amplitudes from $2 \to 2$ scattering of the scalar components (physical and unphysical) of the Higgs doublet (e.g., replace the amplitudes involving scattering of gauge bosons with those involving scattering of Goldstone bosons -- this is an application of the Goldstone boson equivalence theorem).  Include all the coupled channels between net-electrically-neutral two-scalar initial and final states, and work in the high-energy limit.  Express your result in a way that allows you to compare it to the bound found in these lectures (e.g., by using the results of problem 1).

\item A simple way to identify the Goldstone bosons in an extended Higgs sector is to find the linear combination of fields that appears in the (nonphysical) $W^{\pm}_{\mu} \partial^{\mu} \varphi^{\mp}$ and $Z_{\mu} \partial^{\mu} \varphi^0$ ``interaction'' terms coming from the Lagrangian term involving the covariant derivative, which must disappear when the Goldstone bosons are gauged away.  Use this trick to show that the charged Goldstone boson in the scalar septet model is given by
\begin{eqnarray*}
	G^+ = c_7 \phi^+ + s_7 \left( \sqrt{\frac{5}{8}} \chi^{+1} - \sqrt{\frac{3}{8}} (\chi^{-1})^* \right),
\end{eqnarray*}
where $c_7 = v_{\phi}/v$ and $s_7 = 4 v_{\chi}/v$.  The field definitions, covariant derivative, and SU(2) generators for the septet representation can be found in Ref.~\cite{Harris:2017ecz}.

\item Consider the two-Higgs-doublet model with an exact global symmetry $\Phi_1 \to -\Phi_1$ preserved by the scalar potential:
\begin{eqnarray*}
	V &=& m_{11}^2 \Phi_1^{\dagger} \Phi_1 + m_{22}^2 \Phi_2^{\dagger} \Phi_2
		+ \frac{1}{2} \lambda_1 (\Phi_1^{\dagger} \Phi_1)^2 
		+ \frac{1}{2} \lambda_2 (\Phi_2^{\dagger} \Phi_2)^2 \nonumber \\
	&& + \lambda_3 (\Phi_1^{\dagger} \Phi_1) (\Phi_2^{\dagger} \Phi_2)
		+ \lambda_4 (\Phi_1^{\dagger} \Phi_2) (\Phi_2^{\dagger} \Phi_1)
		+ \frac{1}{2} \lambda_5 \left[ (\Phi_1^{\dagger} \Phi_2)^2 + (\Phi_2^{\dagger} \Phi_1)^2 \right],
\end{eqnarray*}
and assume that all the parameters in the scalar potential are real.  
We argued in footnote~\ref{fn:exactZ2} that when both $\Phi_1$ and $\Phi_2$ get vevs as usual, then there is an upper bound on the masses of all the scalars in this model due to perturbativity of the dimensionless couplings $\lambda_i$.  
Consider instead the case that $\Phi_1$ does \emph{not} get a vev (in this case we can set things up so that the fermion masses are all generated by $\Phi_2$ -- this is called the Inert Doublet Model).  Show that, in this situation, there is no such upper bound on the masses of the extra scalars, and that their masses instead grow like $\sqrt{m_{11}^2}$ in the large $m_{11}^2$ limit.

\end{enumerate}

\section*{Acknowledgments}

I thank the Carleton graduate students who took PHYS 6602 in 2017 and 2021 for ``test-driving'' these lectures and my colleague Yue Zhang for coordinating the course in 2021.
This work was supported by the Natural Sciences and Engineering Research 
Council of Canada and by Carleton University, and was performed on the traditional and unceded territory of the Algonquin Anishinabeg people.


\end{document}